%% file: main.tex
\documentclass[letterpaper, 10 pt, conference]{IEEEtran}  % Comment this line out if you need a4paper

\IEEEoverridecommandlockouts                              % This command is only needed if 
\input{preamble}
\title{\LARGE \bf
Privacy-Preserving Stealthy Attack Detection in Multi-Agent Control Systems}

\author{Mohammad Bahrami and Hamidreza Jafarnejadsani% <-this % stops a space
\thanks{Mohammad Bahrami and Hamidreza Jafarnejadsani are with the Department of Mechanical Engineering, Stevens Institute of Technology, Hoboken, NJ, 07030 USA
        {\tt\small \{mbahrami,hjafarne\}@stevens.edu}}%%
}

\begin{document}

\maketitle
\thispagestyle{empty}
\pagestyle{empty}

%%%%%%%%%%%%%%%%%%%%%%%%%%%%%%%%%%%%%%%%%%%%%%%%%%%%%%%%%%%%%%%%%%%%%%%%%%%%%%%%
\begin{abstract}
This paper develops a glocal (global-local) attack detection framework to detect stealthy cyber-physical attacks, namely covert attack and zero-dynamics attack, against a class of multi-agent control systems seeking average consensus. The detection structure consists of a global (central) observer and local observers for the multi-agent system partitioned into clusters. 
The proposed structure addresses the scalability of the approach and the privacy preservation of the multi-agent system's state information. %as it is designed globally and implemented locally.
The former is addressed by using decentralized local observers, and the latter is achieved by imposing unobservability conditions at the global level.
Also, the communication graph model is subject to topology switching, triggered by local observers, allowing for the detection of stealthy attacks by the global observer.
Theoretical conditions are derived for detectability of the stealthy attacks using the proposed detection framework. 
Finally, a numerical simulation is provided to validate the theoretical findings.
\end{abstract}

%%%%%%%%%%%%%%%%%%%%%%%%%%%%%%%%%%%%%%%%%%%%%%%%%%%%%%%%%%%%%%%%%%%%%%%%%%%%%%%%
\section{Introduction}\label{S:introduction}
The grand challenges of ensuring resilience and security in cyber-physical systems (CPS) have motivated the study and characterization of possible adversarial attacks against these complex systems.  Reactive approaches based on the detection and identification algorithms are a significant aspect of comprehensive defense strategies against malicious attacks \cite{cardenas2008secure}. 
Due to their distributed nature, cyber-physical systems such as the power grid or networks of autonomous aerial/ground vehicles can often be modeled as multi-agent systems \cite{pasqualetti2013attack,ren2007distributed}, where the communication network is susceptible to attacks \cite{cardenas2008secure}. 
In particular, this paper considers the problem of detecting stealthy attacks, namely covert attack and zero-dynamics attack, using a scalable detection framework for a class of networked multi-agent systems seeking average consensus upon system's initial conditions, as a canonical cooperative task.

\emph{Literature review:} 
In general, the detection of stealthy attacks is not a trivial problem for networked multi-agent systems. Challenges arise due to the large scale of networked systems and the limited communication capability of its subsystems (or agents), which restrict an effective information aggregation and transmission required to implement centralized approaches \cite{pasqualetti2015divide}. Moreover, the prevalent observer-based attack detectors are ineffective in detecting stealthy attacks, particularly zero-dynamics attack (ZDA) and covert attack that are the worst-case attack scenarios in terms of detectability, due to the fact that they are not observable in the system outputs \cite{pasqualetti2013attack,teixeira2012revealing}.

The conventional detection frameworks for stealthy attacks rely on modifying the system structure or adding redundancy in the system measurements to expose such attacks. For instance, a signal modulation acting on the system actuation to alter the system's input behavior was proposed in \cite{hoehn2016detection} for both covert attack and ZDA detection. Change in the system structure was proposed first in \cite{teixeira2012revealing} upon which the study in \cite{schellenberger2017detection} extended the system dynamics with a randomly switched auxiliary system to achieve non-repeating dynamics, preventing the realization of covert attacks. Most recently, for a class of networks with distinct Laplacian eigenvalues, the authors in \cite{mao2020novel} characterized an intermittent ZDA that remains undetectable regardless of the system's switched structure and obtained the conditions for their detectability. As for the covert attack, the authors in \cite{barboni2020detection} proposed a distributed architecture composed of two cascaded observers for each subsystem to detect the attacks. As another strategy, multi-rate sampling in sampled-data systems was studied in \cite{jafarnejadsani2018multirate,back2017enhancement} to change the direction of sampling zeros and thus to prevent ZDA. Also, distributed function calculation was proposed in \cite{sundaram2010distributed} that requires intensive communication in the network and full knowledge of network model for each node. 

In terms of scalability, considerable effort has been dedicated to extending the existing decentralized and distributed estimation/fault detection methods to the attack detection strategies implementable using locally available information for large-scale systems. For instance, one can refer to secure distributed observers for sensor networks in \cite{mitra2016secure}, distributed attack detection schemes for power networks \cite{pasqualetti2013attack,barboni2020detection,teixeira2010networked,gallo2020distributed}, decentralized detection scheme for stochastic interconnected systems in \cite{anguluri2018attack}, and divide-and-conquer approach in \cite{pasqualetti2015divide}. However, few studies have addressed distributed/decentralized detection strategies for \emph{stealthy} attacks, namely covert attack and zero-dynamics attack \cite{barboni2020detection,gallo2020distributed}. Moreover, they do not address the communication topology switching and the privacy of the agents' information.
 
\emph{Statement of contributions:}
The contributions of this paper are threefold. 
First, as a security objective, we consider the privacy of agents' initial condition and the agreement's final value (consensus) and propose enforced unobservability constraints on the network topology to preserve the network privacy at the global level.
Second, for scalability, we propose a glocal (global-local) attack detection structure for which the networked multi-agent system is partitioned into clusters (subsystems) with their respective globally and locally monitored agents that satisfy specific conditions related to the network privacy and the detectability of stealthy attacks (i.e., zero-dynamics attack and covert attack). 
Finally, we derive the theoretical conditions for topology switching (Theorem \ref{th:Attack_detectability_switching}) under which local detectors trigger switches in the system's communication topology such that stealthy attacks become detectable for the global (centralized) observer. We further discuss different types of topology switching and their outcome for the detection of stealthy attacks.

The rest of the paper is organized as the following. Section \ref{S:problem_formulation} presents the preliminary definitions and the problem formulation. The privacy preserving problem and the attack detection framework are studied in Section \ref{S:privacy_obs}. 
% Section \ref{S:detection} presents the attack detection analysis. 
Section \ref{S:Results} demonstrates the simulation results. Finally, Section \ref{S:conclusion} concludes the paper. 
\section{Problem Formulation}\label{S:problem_formulation}
\subsection{Preliminaries}

\noindent
\textbf{{Notation}.} We use  $ \mathbb{R} $, $ \mathbb{R}_{> 0} $, $ \mathbb{R}_{\geq 0} $, $\cplx$, and $\naturals$ to denote the set of reals, positive reals, non-negative reals, complex, and natural numbers, respectively. Also, $\naturals_{0} = \naturals + \{0\}$.
% We use  $ \mathbb{R} $, $ \mathbb{R}_{> 0} $, $ \mathbb{R}_{\geq 0} $, $ \mathbb{R}^n $ to denote the set of reals, positive reals, non-negative reals, and the $ n $-dimensional Euclidean space, respectively. 
% Sets $\naturals$ and $\cplx$ represent the set of the natural numbers and complex numbers. 
% For any vector $ x \coloneqq \col(x_1,x_2,\ldots,x_n) \in \mathbb{R}^n  $ (matrix $ M \in \mathbb{R}^{m \times n} $), we use the notation $ \lVert \cdot \rVert_p $ to represent the (induced) $ p $-norm $ (1\leq p \leq \infty) $.
We use $ x \coloneqq \col(x_1,x_2,\ldots,x_n) $ to denote (block-partitioned) vectors.
$\boldsymbol{1}_n$, $ \boldsymbol{0}_n$, $ I_n $ and $ 0_{n} $ stand for $n$-vector of all ones, the $n$-vector of all zeros, the identity $ n $-by-$ n $ matrix, and $ n $-by-$ n $ zero matrix, respectively\footnote{We may omit the subscripts when clear from the
context.}. $x^{(\rm m)}(t)$ stands for the
$\rm m$-th order time derivative of $x(t)$. In addition, $ |\cdot| $ denotes the cardinality of sets, and for any index set $\Fc$ with $|{\Fc}|=m$, $I_{\s {\Fc}} \in \real^{n \times m}$ is the concatenation of the $i$-th columns of $I_{n}$ where $i \in {\Fc}$.
%
% We denoted by $\col(x_i)_{i\in \Nc}$ the concatenation of vector- or matrix-valued $x_i$ with component $i$ belong to the set $\Nc$.
%
For a matrix $M \in \real^{m \times n}$,
the range (column space) is defined as $\Img{(M)} = \braces{Mx \mid x \in \real^n} \subseteq \real^m$ and the nullspace is defined as $\ker{(M)} = \braces{x \mid Mx = \boldsymbol{0}} \subseteq \real^n$.
The support of vector $x \in \mathbb{R}^n$ is the set of nonzero components defined as $\supp(x)=\setdef{i\in\braces{1,\dots,n}}{x_i\neq0}$. 
We also define the set of nonzero columns of the $ n $-by-$ n $ matrix $M$ by $\colsupp(M)=\setdef{i\in\braces{1,\dots,n}}{[M]_{:,i}\neq\boldsymbol{0}_{n}}$. 

\noindent
\textbf{{Graph theory}.} Let $ \mathcal{G}=(\mathcal{V},\mathcal{E},\mathcal{A})$ denote a weighted undirected graph with the set of nodes $ \mathcal{V}= \{1,2,\dots,N\} $, set of edges $ \mathcal{E}  \subseteq \mathcal{V} \times \mathcal{V}$, and adjacency matrix $\mathcal{A} \coloneqq  [a_{ij}]\in \mathbb{R}^{\s N}_{\geq0}$.
For any pair of nodes $ i,j,\, i \neq j$, a path from $j$ to $i$ implies the edge $ (i,j) \in \Ec$ corresponding to $a_{ij}>0$, otherwise $a_{ij}=0$. The Laplacian matrix $\mathcal{L} \coloneqq  [l_{ij}] \in 
\real^{n \times n}$ is defined as $ l_{ii} = \sum_{j \neq i}^{}a_{ij}$ and $ l_{ij} = -a_{ij}$ if $i \neq j$. 
By convention, $ \mathcal{N}_i=\{j\in \mathcal{V} \mid \Ec_{ij}\in \mathcal{E} \} $ denotes the set of neighbors of node $ i $. A cluster is defined as any subset $ \boldsymbol{\Pc}:=\{\Pc_1,\dots,\Pc_{\s |\boldsymbol{\Pc}|}\}\subseteq \Vc $ of the nodes of graph $ \mathcal{G} $ such that $ \cup_{i=1}^{|\boldsymbol{\Pc}|}\Pc_i=\mathcal{V} $ and $ \Pc_i\cap \Pc_j=\emptyset $ if $i\neq j$. 
We make the convention that $ \mathcal{G}_{\sigma(t)} $ with a right-continuous switching signal $\sigma(t): \mathbb{R}_{\geq0}\rightarrow{\mathcal{Q}:=\{1,2,\dots,q\}}, \, q\coloneqq|\Qc| $ denotes a finite set of graphs, indexed by finite set $\Qc$, that each holds all properties of graph $\mathcal{G}$. 
\begin{definition}\btitle{Graph component \cite{newman2018networks}}\label{def:graph_component}
A component in an undirected graph is an induced subgraph with a (maximal) subset of nodes such that each is reachable by some path from each of the others.
% and such that no other node in the graph can be added to the subset while preserving this property.
% % another version:
% % in an undirected graph, a component to which a node belongs is a (maximal) subset of nodes that can be reached from it by paths running along edges of the graph, and such that no other node in the graph can be added to the subset while preserving this property.
\end{definition}

\noindent
\textbf{{Systems theory.}} A linear system $ \dot{x}(t)=Ax(t)+Bu(t)$, $ y(t)=Cx(t)+Du(t) $, where $x(t) \in \real^n, u(t) \in \real^m, y(t)\in \real^p$, is represented by the tuple $\Sigma(A,B,C,D)$.
\begin{definition}[Zeroing direction and zero-dynamics attack {\cite[Ch. 3]{zhou1996robust}},\cite{mao2020novel}]\label{def:inv_zeros}
Scalar $ \lambda_0\in \mathbb{C} $ is a zero of the tuple $\Sigma(A,B,C,D)$ if, and only if, there exists zeroing direction $\col(\mb{x}_0,\mb{u}_0)\neq \col(\boldsymbol{0},\boldsymbol{0})$ associated with $ \lambda_0 $ such that 
\begin{equation}\label{eq:inv_zeros}
    \begin{bmatrix}
        \lambda_0 I_n - A & -B \\ C & D
    \end{bmatrix}
    \begin{bmatrix}
        \mb{x}_0  \\ \mb{u}_0
    \end{bmatrix}
    =
    \begin{bmatrix}
        \boldsymbol{0} \\ \boldsymbol{0}
    \end{bmatrix}.
\end{equation}
Then, the signal $ u(t)=\mb{u}_0e^{\lambda_0t}$ is a zero-dynamics attack that generates non-zero state trajectories $x(t)=\mb{x}_0e^{\lambda_0t}$ while the output $ y =Cx+Du$ satisfies $y(t) =\boldsymbol{0}$. 
\end{definition}
%%
%%
%%%%%%%%%%%%%%%%%%%%%%%%%%%%%%%%%%%%%%%%%%%%%%%%%%%%%%%%%%%%%%%%%%%%%%%%%%%%%%%%
\subsection{Problem Statement}
\noindent
\textbf{{System model.}} Consider a graph $\mathcal{G}$ of order $ N $, we associate each node $i$ of the graph with an agent $ \Sigma_i $ that evolves according to the following dynamics\footnote{For brevity, we may omit the time
argument $t$ from expressions whenever possible in the rest of the paper.}:
\begin{equation}\label{eq:ol_sys}
\Sigma_{i}:
\left\{
\begin{array}{l}
\dot{x}_i(t) = v_i(t)\\ 
\dot{v}_i(t) = u_i(t) 
\end{array},
\right. \qquad i\in \mathcal{V}, \\
\end{equation}
in which $ x_i(t) $ and $ v_i(t) $ denote the position and velocity, and $u_i(t) $ (to be determined) stands for the control channel through which each agent communicates with a set of neighbors $\mathcal{N}_i $ to perform a prespecified cooperative task.

\noindent
\textbf{{Control protocol}.} The objective is to reach an average consensus upon the initial conditions of the system, as follows:
\begin{equation}\label{eq:cond_consensus}
\lim \limits_{t\rightarrow{\infty}} \left|x_i(t)-x_j(t) \right| =0
\; \text{and} \;
\lim \limits_{t\rightarrow{\infty}} \left|v_i(t) \right| =0, \quad \forall \; i,j \in \mathcal{V},
\end{equation}
which can be achieved by exchanging local information through the following  switching control protocol \cite{mao2020novel}:
\begin{align}\label{eq:ctrl_proto}
    u_i &=  - \gamma v_i -\alpha \sum_{j\in \mathcal{N}_i} a^{\sigma(t)}_{ij}(x_i-x_j)  + u_{a_i}, \quad i \in \Vc,
\end{align}
where $ a^{\sigma(t)}_{ij} $ is the entry of the symmetric adjacency matrix associated with the graph $ \mathcal{G}_{\sigma(t)} $
representing the switching communication network of agents $ \Sigma_i $'s. Also, $\alpha$ and $\gamma$ are the control gains. Finally, $u_{a_i} $ is the injected malicious signal in control channel of the $i$-th agent. We assume the unknown subset $\overline{\Fc} \subset \Vc$ represents the set of compromised agents, and we have $u_{a_i} =0 $ for an uncompromised agent $i$, i.e., if $ i \in \Vc \setminus \overline{\Fc} $.
%
%=========== add switching definition here 
%

\noindent
\textbf{{Closed-loop system.}} Given \eqref{eq:ol_sys} and \eqref{eq:cond_consensus}, let $ \mb{x} \coloneqq \col(x,v)$, where $x \coloneqq \col(x_1,\dots,x_{\s N})$, $v \coloneqq \col(v_1,\dots,v_{\s N})$, and $\mb{u}_a=\col(u_{a_i}),\, i \in  \overline{\Fc}$. Then, the closed-loop system is given by
\begin{equation}\label{eq:cl_sys}
\Sigma:
\left\{
\begin{array}{l}
\underbrace{
    \begin{bmatrix}
    \dot{x} \\ \dot{v}
    \end{bmatrix}
    }_{\dot{\mb{x}}}
    =
    \underbrace{
    \begin{bmatrix}
    0_{} & I_{} \\
    -\alpha \mathcal{L}_{\sigma(t)} &  -\gamma I_{}
    \end{bmatrix}
    }_{\mb{A}_{\sigma(t)}}
    \underbrace{
     \begin{bmatrix}
    x \\ v
    \end{bmatrix}
    }_{\mb{x}}
    + \underbrace{
    \begin{bmatrix}
    0_{\s } \\ I_{\s \overline{\Fc}}
    \end{bmatrix}
    }_{\mb{B}}\mb{u}_a, \\ 
    \mb{x}(t_0)=\mb{x}_0,\\
    \mb{y} ={\mb{C}}{\mb{x}}-\mb{u}_s, \ \ \mb{C}=\diag\braces{C_{\rm x}, C_{\rm v}}, 
\end{array}
\right.  
\end{equation}
with the system measurements $\mb{y}=\col(\mb{y}_1,\cdots,\mb{y}_{\s |\Mc|})$ corresponding to the output matrix $\mb{C}$ such that:
\begin{equation}\label{eq:measurments}
  \colsupp({C_{\rm k}}) \in \Mc_{\rm k} \subset \Vc, \ \ {\rm k} \in \braces{{\rm x},{\rm v}}, \ \ \Mc =  \braces{\Mc_{\rm x}, \Mc_{\rm v}},
\end{equation}
where the to-be-selected set $\Mc$ represents the set of the monitored agents' index. 
Also, $\mb{u}_s=\col(u_{s_1},\dots,u_{s_{\s|\Mc|}})$ is a vector of injected malicious signals in the compromised measurement sensor channels. Finally, the Laplacian $ \mathcal{L}_{\sigma(t)} $ in \eqref{eq:cl_sys} encodes the information exchange among agents.

\noindent
\textbf{{Adversary model.}}
%
% \MB{$\Fc_u\cap \Fc_s=\emptyset?$}
Let $ \overline{\Fc} \subset \Vc $ denote the set of agents with a compromised (under attack) control channel, and 
$ \underline{\Fc} \subset \Mc $ represent the set of agents with compromised sensor channels. The dynamics of the adversarial attack is given by\footnote{The matrix $\mb{B}$ in \eqref{eq:attack_model} is the same as in \eqref{eq:cl_sys}. it is designed by the attacker.}
\begin{equation}\label{eq:attack_model}
\Sigma_{\s \Ac}:
\left\{
\begin{array}{l}
\dot{\tilde{\mb{x}}}=\tilde{\mb{A}}_{{\sigma}(t)}\tilde{\mb{x}}
+{\mb{B}}\mb{u}_a(t), 
\quad \tilde{\mb{x}}(t_a)=\tilde{\mb{x}}_0, \\ 
\mb{u}_s =\tilde{\mb{C}}_{\s}\tilde{\mb{x}},\\
\supp(\mb{u}_a)=\overline{\Fc}, \quad \supp(\mb{u}_s)=\underline{\Fc},
\end{array}
\right.  
\end{equation}
where the vector attack $\mb{u}_a$ is generally a function of disclosed information, i.e., $\mb{u}_a \coloneqq f(\tilde{\mb{x}},u_i,\mb{y}_{},t)$ by which the attacker steers the system towards undesired states, and $t_a \ge t_0$ is the attack starting time. 
% Here we focus on the case $\mb{u}_a \coloneqq f(t)$.
For example, the attack signal is in the form of $\mb{u}_a(t)=\mb{u}_0e^{\lambda_0(t-t_a)}$ in the case of ZDA, where $\lambda_0$ and $\mb{u}_0$ are introduced in Definition \ref{def:inv_zeros}. 

\noindent
\textbf{{Communication topology switching}.} The multi-agent system in \eqref{eq:cl_sys} operates in the \textit{normal mode} with the initial communication topology specified by $\sigt=1\in \Qc,$ $ t\in [t_0, t_1)$ until switching to a \textit{safe mode} following the detection of an attack at the time $t_1 > t_a$. In the safe mode for $t \ge t_1$, the communication topology switching is specified by the switching signal $\sigt=\braces{2,\dots,q}\in \Qc$, $q\coloneqq|\Qc|$ whose switching policy will be determined later (See Section \ref{S:detection}).
\begin{assumption}\btitle{Disclosed information}\label{assum:attacker} In the normal mode, where $\sigt=1\in \Qc,$ $ t\in [t_0, t_1)$, the attacker
\begin{enumerate}
    \item has perfect knowledge of the system model, i.e., $\Sigma_{\s \Ac}(\tilde{\mb{A}}_{ \sigma(t)},{\mb{B}},\tilde{\mb{C}}_{},\sigma=1)
    =
    \Sigma({\mb{A}}_{ \sigma(t)},\mb{B},\mb{C}_{},\sigma=1)$, 
    % before any switches in system structure (communication topology),
    \item does not know the system's initial condition, i.e., $ \tilde{\mb{x}}(t_a) \neq {\mb{x}}(t_0) $, and $ \tilde{\mb{x}}(t_a) = \tilde{\mb{x}}_0 = \boldsymbol{0} $ in a covert attack.
    % $ \tilde{\mb{x}}(t_a) =  \boldsymbol{0}  \Rightarrow  {\mb{x}}(t_a) = \bar{\mb{x}}(t_a)\!$ in a covert attack.
    \item has no knowledge of the system switching times $\braces{{t}_k}_{k=1}^{\mb{m}-1}$, $\mb{m}\in \naturals$ associated with the safe mode when  $\sigt=\braces{2,\dots,q}\in \Qc,$ $ t\in [t_1, +\infty)$, 
    \item starts the attack at $t_a\geq t_0=0$.
    % has no knowledge of the system switching times $\braces{{t}_k}_{k=1}^{q-1}$ that activate the backup modes of the communication topology associated with $\sigt=\braces{2,\dots,q}\in \Qc,$ $ t\in [t_1, +\infty)$, 
    % \item[(\emph{iv})] starts the attack at $t_a\geq t_0$.
\end{enumerate}
\end{assumption}
\begin{assumption}\btitle{Defender's policy}\label{assum:defender}
The defender
\begin{enumerate}
    \item selects the monitored agents and designs the attack detection framework,
    \item designs the communication topology for the safe mode and its corresponding switching policy.
\end{enumerate}
\end{assumption}
For the detectability of adversarial attacks in switched systems, we will need the following technical result:
\begin{lemma}\btitle{Observability of linear switched systems \cite{tanwani2012observability}}\label{lemma:observability}
Given a system $\dot{\mb{x}}=\mb{A}_{\s \sigt}\mb{x}$, with measurements  $\mb{y}_{}=\mb{C}\mb{x}$, ($\mb{x} \in \real^n$ and $\mb{y} \in \real^p$), over the interval $t\in[t_{0},{t}_{\mb{m}})$ that includes switching instances $\braces{{t}_k}_{k=1}^{\mb{m}-1}$ for modes $\sigt=k\in\Qc$ with the dwell time ${\tau}_{k}={t}_{k}-{t}_{k-1}$, the output of system is given by $\mb{y}(t)=\mb{C}e^{\mb{A}_k(t-t_{k-1})}\prod_{l=k-1}^{1}e^{\mb{A}_l(\tau_{l})}\mb{x}(t_0), t\in[t_{k-1}, \, t_{k})$.
Then, (i) the system is observable and the initial condition $\mb{x}(t_0)$ is reconstructable from $\mb{y}(t)$ if, and only if, \eqref{eq:obsv_O} is full rank (i.e., $\boldsymbol{\Nc}_1^{\mb{m}} \coloneqq \ker(\boldsymbol{\Oc})=\{0\}$).
(ii) If \eqref{eq:obsv_O} is rank deficient, the unobservable subspace of the system for $t\in[t_{0},t_{\mb{m}})$, which is the largest $\mb{A}_{\s \sigt}$-invariant subspace contained in $\ker(\mb{C})$, can be recursively computed using \eqref{eq:obsv_nullspace_1}-\eqref{eq:obsv_nullspace_2}. 
% %
\begin{align}
% %
\label{eq:obsv_O}
\boldsymbol{\Oc}&=\col(\Oc_1,\Oc_2 e^{\mb{A}_1\tau_1},\cdots,\Oc_{\mb{m}}\prod_{i=\mb{m}}^{1}e^{\mb{A}_{i}\tau_{i}}),\\
\label{eq:obsv_nullspace_1}
\boldsymbol{\Nc}_{\mb{m}}^{\mb{m}} &= \ker(\Oc_{\mb{m}}),\\
% %
\label{eq:obsv_nullspace_2}
\boldsymbol{\Nc}_{k}^{\mb{m}} &= \ker(\Oc_{k})
\cap \bracket{\bigcap_{i=k+1}^{\mb{m}} \ker\paren{\Oc_{i} \prod_{j=i-1}^{k}e^{\mb{A}_{j}\tau_{j}}}}, \\ \nonumber
% %
\text{where}\\
\label{eq:obsv_nullspace_matrix}
\Oc_{k} &= \col\paren{\mb{C},\mb{C}\mb{A}_{k},\dots,\mb{C}\mb{A}_{k}^{\s 2N-1}}, \,\, 1\leq k \leq \mb{m}\!-\!1,\\
% %
\label{eq:obsv_nullspace_Amatrix}
\mb{A}_k &= \mb{A}_{\sigma(t)}, \quad t\in [t_{k-1},t_{k}).
%%%
% %
\end{align}
% %
\end{lemma}

\begin{proposition}\btitle{Stealthy attacks}\label{propo:stealthy_attacks}
Consider system \eqref{eq:cl_sys}, under the attack model \eqref{eq:attack_model} and Assumption \ref{assum:attacker}, an attack is stealthy\footnote{The stealthy attacks defined by the condition \eqref{eq:stealthy} are also known as undetectable attacks in the literature \cite{pasqualetti2013attack}.} if the system output in \eqref{eq:cl_sys} satisfies  
\begin{equation}\label{eq:stealthy}
   \mb{y}_{}(\mb{x}_0,\mb{u}_a,\mb{u}_s,t)=\mb{y}_{}(\bar{\mb{x}}_0,\mb{0},\mb{0},t),\quad \forall \, t\in [t_0,t_1), 
\end{equation}
where $\mb{x}_0$ and $\bar{\mb{x}}_0$ are the actual and possible initial states, respectively. Then, \eqref{eq:stealthy} can be realized in two senses 
\begin{enumerate}
    \item {Covert Attack}: 
    Under Assumption \ref{assum:attacker}, if the attacker sets the initial condition $\tilde{\mb x}(t_a)=\boldsymbol{0}$ or alternatively $\tilde{\mb x}(t_0)\in \boldsymbol{\Nc}_1^{1}=\ker(\Oc_1)$ in \eqref{eq:attack_model}, then the attack $\mb{u}_a$ on \eqref{eq:cl_sys} is covert, that is there exists a vector $\mb{u}_s$, injected in \eqref{eq:cl_sys}, canceling out the effect of $\mb{u}_a$ on the system output $\mb{y}(t)$.
    \item {Zero-dynamics Attack (ZDA)}: the attacker can excite the zero dynamics of the system by an unbounded signal and remains stealthy with no need to alter the system measurements (i.e., $\mb{u}_s(t)=\boldsymbol{0}$ in \eqref{eq:cl_sys}) if $\tilde{\mb{x}}_0 \in \ker(\mb{C}_{})$ and $\mb{u}_a(t)=\mb{u}_0e^{ \lambda_0(t-t_a)},\, t_a=t_0$, where $\lambda_0$, $\tilde{\mb{x}}_0$ and $\mb{u}_0$ are obtained using Definition \ref{def:inv_zeros}.
\end{enumerate}
\end{proposition}
\begin{IEEEproof}
Clearly before an attack starts, \eqref{eq:stealthy} is met over $t\in [t_0,t_a)$. Consider $\mb{x}(t_a) $ as the system states when the attack starts,\\
({\textrm{i}}): in the case of covert attack, the output of the system \eqref{eq:cl_sys} with the initial normal mode $\sigt=1$ over $t\in[t_a,t_{1})$ is given by
{\small
\begin{align}\label{eq:ym_solution}
	\mb{y}_{}(t) &= 
	\mb{C}_{} 
	{e^{\mb{A}_1(t-t_{a})}}
	\mb{x}(t_a) 
	+
	\mb{C}_{}
	{\int^{t}_{t_{a}}
	e^{\mb{A}_{1}(t-\boldsymbol{\tau})}
	\mb{B}\mb{u}_{a}(\boldsymbol{\tau})d{\boldsymbol{\tau}}}
    - \mb{u}_s(t),
% 	\nonumber\\
% 	&\quad\:{-}\:\mb{u}_s(t),
\end{align}
}
and the last term which is the output of the attacker's model \eqref{eq:attack_model} is given by
{\small
\begin{align}\label{eq:us_solution}
	\mb{u}_{s}(t) &=
	\tilde{\mb{C}}_{} 
	{e^{\tilde{\mb{A}}_1(t-t_{a})}}
	\tilde{\mb{x}}(t_a)
	+
	\tilde{\mb{C}}
	{\int^{t}_{t_{a}}
	e^{\tilde{\mb{A}}_1(t-\boldsymbol{\tau})}
	\mb{B}\mb{u}_{a}(\boldsymbol{\tau})d{\boldsymbol{\tau}}}.
\end{align}
}
Substituting \eqref{eq:us_solution} into \eqref{eq:ym_solution} and considering Assumption \ref{assum:attacker} yields 
	\begin{eqnarray}\label{eq:ym_solution_final}
	\mb{y}_{}(t) =
	\mb{C}_{} {e^{\mb{A}_1(t-t_{a})}}(\mb{x}(t_a)-\tilde{\mb{x}}(t_a)), \quad t\in[t_a,t_{1}).
	\end{eqnarray}
The measurement \eqref{eq:ym_solution_final} matches the attack-free response if the attacker simply sets $\tilde{\mb x}(t_a)=\boldsymbol{0}$. Also, in the case $\tilde{\mb x}(t_a)\neq \boldsymbol{0},\, t_a=t_0=0$, it is immediate from lemma \ref{lemma:observability} that if $\tilde{\mb x}(t_0)\in \boldsymbol{\Nc}_1^{1}\neq\braces{0} \implies \mb{C}_{} {e^{\mb{A}_1(t-t_{a})}}\tilde{\mb{x}}(t_0)=\boldsymbol{0},\, t_a=t_0=0$ in \eqref{eq:ym_solution_final}, and thus $\mb{y}(t) = $ $ \mb{C}_{} {e^{\mb{A}_1(t-t_{a})}}\mb{x}(t_a),$ $ t \in [t_a,t_{1})$. In both of the cases, condition \eqref{eq:stealthy}, guaranteeing the covertness of the attack, is met. We, however, focus on the first case under Assumption \ref{assum:attacker}-{\rm (ii)}, therefore the system state $\mb{x}(t)$, without any jump, continuously holds the following
%
%===========
% there is a third case that in (16): x-\tilde{x} := \bar{x} and y satisfies (13)??
%===========
%
\begin{align}\label{eq:covert_trajec}
  \mb{x}(t)&=\bar{\mb{x}}(t)+\tilde{\mb{x}}(t),
 \end{align}
  where 
  \begin{align}\label{eq:covert_trajec_pre}
  \tilde{\mb{x}}(t)&=\boldsymbol{0} 
  \implies
  \mb{x}(t) = \bar{\mb{x}}(t), \ \  &\forall\, t \in [t_0, t_a),
  \\ \label{eq:covert_trajec_post}
  \tilde{\mb{x}}(t)&= 
	{\int^{t}_{t_{a}}
	e^{{\mb{A}}_{1}(t-\boldsymbol{\tau})}
	\mb{B}\mb{u}_{a}(\boldsymbol{\tau})d{\boldsymbol{\tau}}}, 
	\ \  &\forall\, t \in [t_a, t_1),
\end{align}
with $\bar{\mb{x}}(t),\, \forall\, t \in [t_0, t_1)$ denoting the state of the system in \eqref{eq:cl_sys} in the absence of covert attack (i.e. $\dot{\bar{\mb{x}}}=\mb{A}_{1}\bar{\mb{x}},\, \bar{\mb{x}}_0={\mb{x}}_0$).
% This implies there is no jump in system state $\mb{x}(t)$.
\\
(\textrm{ii}): In the case of ZDA, let $t_a=t_0=0$ for simplicity, and $\bar{\mb{x}}_0 = \mb{x}_0 -\tilde{\mb{x}}_0$. Under Assumption \ref{assum:attacker} and using Definition \ref{def:inv_zeros}, the attacker can solve the following:
\begin{equation}\label{eq:zda_cond}
    \begin{bmatrix}
        \lambda_{\s 0} I_{} -\mb{A}_{1} & -\mb{B} \\ \mb{C}_{} & 0_{}
    \end{bmatrix}
    \begin{bmatrix}
        \tilde{\mb{x}}_0  \\ \mb{u}_0
    \end{bmatrix}
    =
    \begin{bmatrix}
       \boldsymbol{0}  \\ \boldsymbol{0}
    \end{bmatrix},
\end{equation}
to design the ZDA signal $\mb{u}_a(t)=\mb{u}_0e^{ \lambda_0 t}$ causing unbounded system states
\begin{align}\label{eq:zda_trajec}
    \mb{x}(t)=\bar{\mb{x}}(t)+\tilde{\mb{x}}_0e^{ \lambda_0 t},
\end{align} 
while \eqref{eq:stealthy} is met, where $\bar x(t)$ is the state of the system in (5) assuming the initial condition $\bar x_0$ and no attack signal.
The second equation in \eqref{eq:zda_cond}, $\mb{C}\tilde{\mb{x}}_0=\boldsymbol{0}$, implies $\tilde{\mb{x}}_0\in \ker(\mb{C}_{})$. 
It is an immediate result from Definition \ref{def:inv_zeros} that the attack signal $\mb{u}_a(t)=\mb{u}_0e^{\lambda_{\s 0}t}$ results in $\mb{u}_s(t)=\mb{C}\tilde{\mb{x}}(t)=\boldsymbol{0}$ in \eqref{eq:attack_model} while the system states $\tilde{\mb{x}}(t)=\tilde{\mb{x}}_0e^{\lambda_0t} \in \ker(\mb{C}), \, \forall \,t\in[t_0,t_{1})$ is unboundedly increasing. 
% Recall $\tilde{\mb{x}}_0 = \mb{x}_0- \bar{\mb{x}}_0$ and the superposition principle in linear systems, 
Consider \eqref{eq:zda_trajec} and the superposition principle in linear systems, 
then injecting the designed ZDA signal $\mb{u}_a(t)$ in \eqref{eq:cl_sys} yields the solution $\mb{y}=\mb{C}\mb{x}(t)=\mb{C}\bar{\mb{x}}(t)+\mb{C}\tilde{\mb{x}}_0e^{ \lambda_0 t}$, which by considering \eqref{eq:zda_cond} is equivalent to \eqref{eq:stealthy}, guaranteeing the stealthiness of ZDA for \eqref{eq:cl_sys}.
\end{IEEEproof}

Given the system and attack models above, we now state the two problems which this paper aims to address in the following:

\begin{problem}\btitle{Privacy-preserving average consensus}\label{prob:privacy}
% Given the switched system \eqref{eq:cl_sys} subject to consensus constraint \eqref{eq:cond_consensus},
Given the switching consensus system \eqref{eq:cl_sys},
we seek to preserve the following privacy requirements:
\begin{enumerate}
    \item \label{privacy_requr_IC} neither system's initial states $\mb{x}(t_0)$ nor final agreement values ($x^*=\frac{1}{N}\sum_{i=1}^Nx_i(t_0)$, $v^*=0$) should be revealed or be reconstructable.
    \item \label{privacy_requr_topol_reconst} the system's communication topology $\Gc_{\sigt}$ should not be reconstructable.
\end{enumerate}
\end{problem}
%
%noise-based obfuscation%
\begin{problem}\btitle{Scalable attack detection}\label{prob:attack_detection}
Given the system in \eqref{eq:cl_sys} under the attack model \eqref{eq:attack_model}, we seek to develop a stealthy attack detection framework such that:
\begin{enumerate}
    \item it features a decentralized and scalable structure.
    \item it satisfies the privacy-preserving requirements defined in Problem \ref{prob:privacy}.
\end{enumerate}
\end{problem}
%%
%%%%%%%%%%%%%%%%%%%%%%%%%%%%%%%%%%%%%%%%%%%%%%%%%%%%%%%%%%%%%%%%%%%%%%%%%%%%%%%%
\section{Privacy Preservation and Attack Detection}\label{S:privacy_obs}
In this section, we describe the attack detection framework and characterize the conditions required to address Problems \ref{prob:privacy} and \ref{prob:attack_detection}. 
\subsection{Attack Detection Scheme}\label{subS:attak_detection_scheme}
\begin{figure}[t]
    \centering
    \includegraphics[width=.9\linewidth]{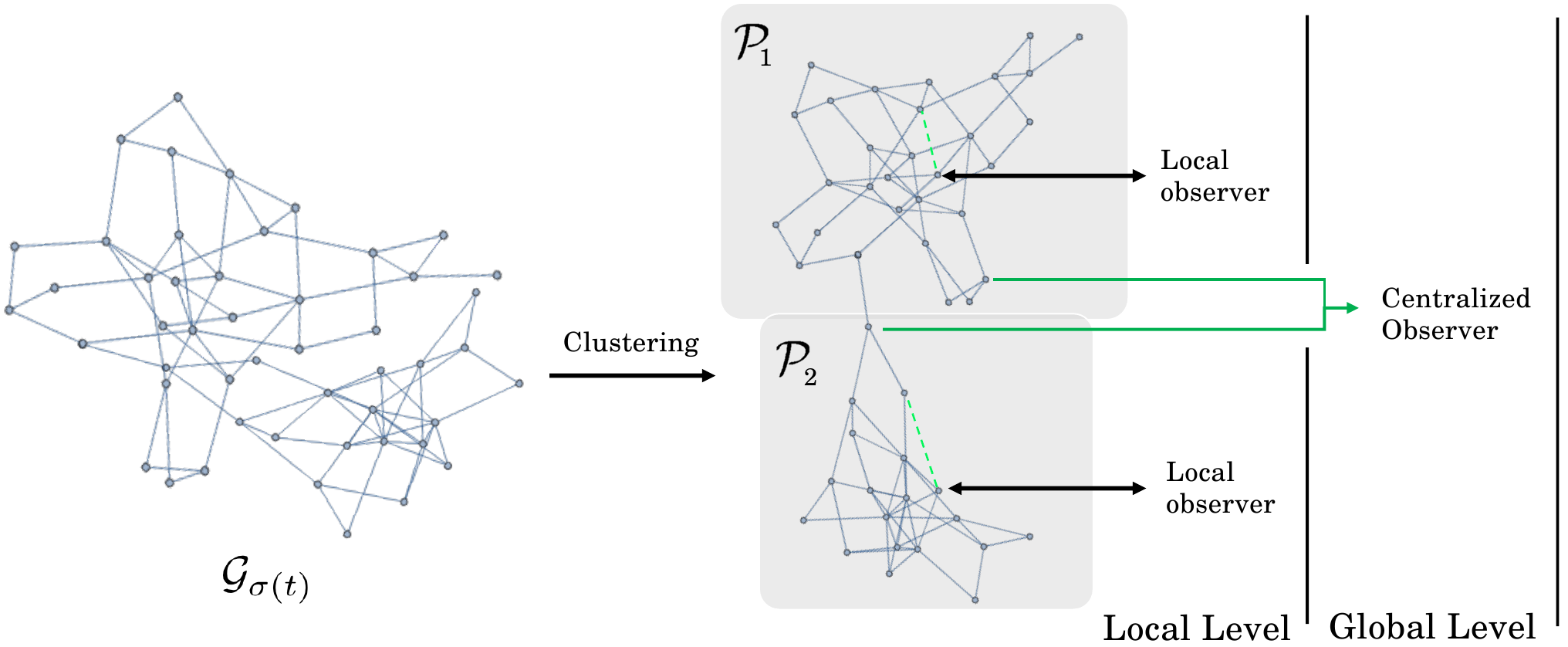}
    \caption{\small Attack detection architecture.} 
    \label{fig:detection_scheme}
\end{figure}
The proposed framework, depicted in Figure \ref{fig:detection_scheme}, is a two-level attack detection framework. It is privacy-preserving and relies on topology switching generating model discrepancy between the attacker model \eqref{eq:attack_model} and the actual system \eqref{eq:cl_sys}. 
The system is decomposed into a set of subsystems based on the characteristics of its communication topology such as sparsity. Then, a set of monitored agents will be characterized such that each subsystem (the dynamics of agents within a cluster) is fully observable with respect to its locally available measurements while the main system \eqref{eq:cl_sys} is partially observable with respect to its globally available measurements \eqref{eq:measurments}.
We show how unobservability and system clustering can be used respectively to address Problem \ref{prob:privacy} and \ref{prob:attack_detection}.
Building upon global and (private) local measurements, the attack detection framework consists of a centralized observer, implemented in the control center, and local observer(s) in each cluster ($ \Pc_{\rm i} ,\, {\rm i}\in \{1,2\}$ in Figure \ref{fig:detection_scheme}). As increasing data transmission between agents and the centralized observer in the control center arises scalability and privacy concerns (cf. Problem \ref{prob:attack_detection}),
local observers play a vital role in our attack detection framework. They are hidden from the attacker because they are distributed among clusters of the multi-agent system, and their output is not sent to the control center but kept locally for attack detection.
If a local observer detects a stealthy attack, it triggers a network topology switch whereby the stealthy attack becomes detectable in the global measurements available for the centralized observer. The local decision making for network topology switches and indirect communication with the control center allow for agile reconfigurability in autonomous multi-agent systems (e.g. a network of autonomous aerial/ground vehicles) as well as eliminates the need for additional data exchange required, at global level, for monitoring and stealthy attack detection.

\subsection{Privacy Preservation}\label{Ss:privacy}
Problem \ref{prob:privacy} on privacy preservation can be addressed by imposing unobservability constraint on system \eqref{eq:cl_sys}. Indeed, one can select the set of monitored agents $\Mc$ in \eqref{eq:measurments} such that $(\mb{A}_{\sigt},\mb{C})$ is not an observable pair on $t\in[t_0,+\infty)$, making the globally available measurement $\mb y$ in \eqref{eq:measurments} insufficient to reconstruct either the entire system states' information or the system's switching structure (cf. privacy requirements in Problem \ref{prob:privacy}). 

The following lemma provides sufficient conditions to determine whether the global system measurement \eqref{eq:measurments} is consistent with the privacy requirements.
% %
\begin{lemma}\btitle{Invariant unobservable subspace of  system \eqref{eq:cl_sys}}\label{lemma:Inv_unobs_subspace}
The subspace $ \spann \braces{\begin{smallmatrix} \boldsymbol{1}_{\! N} \\ \boldsymbol{0}_{\!N} \end{smallmatrix}}$ is an $\mb{A}_{\sigt}$-invariant unobservable subspace of the switching system in \eqref{eq:cl_sys} provided that 
it lies in $\ker(\mb{C}) $ and $\Gc_{\sigt}$ features only connected undirected (or strongly connected and balanced directed) graphs.
\end{lemma}
\begin{IEEEproof} See Appendix \ref{app_Inv_unobs_subspace}.
\end{IEEEproof}
\begin{remark}\btitle{Generality of Lemma \ref{lemma:Inv_unobs_subspace}} The result suggests that monitoring only the agents' velocity causes the agents' positions not to be reconstructable independently for system \eqref{eq:cl_sys}. This is a generic solution to Problem \ref{prob:privacy} that holds for all undirected graphs. It is also worth noting that the monitored agents corresponding to set $\Mc$ in \eqref{eq:measurments} can be also selected differently from the results in Lemma \ref{lemma:Inv_unobs_subspace} for any particular graph.
\end{remark}

We next introduce the system partitioning method followed by observer design to address Problem \ref{prob:attack_detection}.
\subsection{System Partitioning}
Consider the communication graph $\Gc_{\sigt}=(\Vc,\Ec,\Ac)$ of the system \eqref{eq:cl_sys}, let the set of agents $\Vc$ be partitioned into disjoint clusters $\boldsymbol{\Pc} \coloneqq \{\Pc_1,\dots,\Pc_{\s |\boldsymbol{\Pc}|}\} $ such that $\cup_{\rm{i}=1}^{\s |\boldsymbol{\Pc}|}\Pc_{\rm{i}}=\Vc$ with $\Pc_{\rm{i}} \in \real^{N_{\rm{i}}}$ and inter-cluster couplings 
 $\Ec_{\rm{cut}} \coloneqq \setdef{\Ec_{{ij}}}{ {i}\in \Pc_{\rm i},\, {j}\in \Pc_{\rm j},  \, \Pc_{\rm i} \cap \Pc_{\rm j} = \emptyset}$. 
Accordingly, after relabeling the system states, the system \eqref{eq:cl_sys} is partitioned into $|\boldsymbol{\Pc}|$ subsystems described as
\begin{equation}\label{eq:cluster_dyn2}
\Sigma_{\scriptscriptstyle \mathcal{P}_{{\rm i}}}:
\left\{
  \begin{aligned}
\dot{\mb{x}}_{\rm{i}} &= \mb{A}_{ \sigt}^{{\rm{i}}}\mb{x}_{\rm{i}} 
+
\sum\nolimits_{\rm{j}\in \Nc_{\spi}} \mb{A}_{{\sigt}}^{\rm{ij}} \mb{x}_{\rm{j}} + \mb{B}^{{\rm i}} \mb{u}_{a_{\s \rm{i}}},  \\
\mb{y}_{{\rm i}_i} &= 
\mb{C}_{{\rm i}_i}\mb{x}_{{\rm i}}, \quad i \in \Mc_{{\rm i}} \subset \Pc_{{\rm i}}, \\
\mb{x}_{{\rm i}}(0)&=
\mb{x}_{0_{{\rm i}}},  \qquad  \, {\rm i} \in
\braces{1,\cdots,{|\boldsymbol{\Pc}|}},
  \end{aligned}\right.
\end{equation}
with
\begin{align}
    \mb{A}_{\sigt}^{\rm{i}}  &= 
    \begin{bmatrix}
    0_{} & I_{} \\
    -\alpha \mathcal{L}^{\rm}_{\sigma(t)} &  -\gamma I_{}
    \end{bmatrix}, \:
    \mb{A}_{\sigt}^{\rm{ij}}  = 
    \begin{bmatrix}
    0_{} & 0_{} \\
    -\alpha \mathcal{L}^{\rm{ij}}_{\sigma(t)} &  0_{}
    \end{bmatrix},\\
    \Lc_{\sigt} &= \begin{bmatrix} 
  \mathcal{L}^{\s 1}_{\sigma(t)} & \cdots &  \mathcal{L}^{\s1,|\boldsymbol{\Pc}|}_{\sigma(t)} \\
  \vdots & \ddots & \vdots\\
    \mathcal{L}^{\s|\boldsymbol{\Pc}|,1}_{\sigma(t)} & \cdots &  \mathcal{L}^{\s|\boldsymbol{\Pc}|}_{\sigma(t)}
    \end{bmatrix}, \,
    \mb{B}^{\rm i}= \begin{bmatrix} 
     0_{\s }  \\ I_{\s \overline{\Fc}_{\rm i}}
    \end{bmatrix},
\end{align}
where $\mb{x}_{ \rm i} \coloneqq \begin{bmatrix}(x)_{\rm i}^{\top} & (v)_{\rm i}^{\top} \end{bmatrix}^{\top} \! \in  \real^{2 N_{{\rm i}}}$ with $(x)_{\rm i}$ and $(v)_i$ representing the vectors of position and velocity states belonging to cluster $\Pc_{\rm i} \subset \Vc$. 
% %
    Also, $\mb{u}_{a_{\s \rm i}}$ associated with the set $\overline\Fc_{\rm i}$ is the vector-valued attack on actuator channels in the cluster as defined in \eqref{eq:cl_sys}.
    % and
    % $\bar{\mb{u}}_{s_{\s i}}$ is a vector of possible adversaries on the local measurements $y_i$. 
    The output signal $\mb{y}_{{\rm i}_i}(t)$, associated with the output matrix $\mb{C}_{{\rm i}_i}$, denotes the \emph{local} measurements that are available at node $i$ in cluster $\Pc_{\rm i}$.
    Finally, $ \mathcal{N}_{\scriptscriptstyle \mathcal{P}_{{\rm i}}}:=\{{\rm j} \in \braces{1,\cdots,{|\boldsymbol{\Pc}|}} \mid \exists \, \Ec_{i,j} \in  \Ec_{\rm{cut}}, i \in \Pc_{\rm i}, j \in \Pc_{\rm j} \} $ denotes the index set of the neighboring clusters of cluster $ \Pc_{\rm i} $.
    
    We note that the decomposition of \eqref{eq:cl_sys} into \eqref{eq:cluster_dyn2} leads to a concatenated set $ \bar\Mc \coloneqq \braces{\Mc,\Mc_1,\dots,\Mc_{|\boldsymbol{\Pc}|}} $, where the set $\Mc$ is associated with \emph{global} measurements \eqref{eq:measurments} available for the control center and sets $\Mc_{\rm i}\text{'s}, \; {\rm i} \in \boldsymbol{\Pc}$ are associated with the \emph{local} measurements $\mb{y}_{{\rm i}_i}$ available at a node $i$ in respective clusters $\Pc_1,\dots, \Pc_{\s |\boldsymbol{\Pc}|}$ in \eqref{eq:cluster_dyn2}.

    We make the following assumptions:
    \begin{assumption}\btitle{Local information}\label{assum:local_info}
\begin{enumerate}
    \item \label{assum:local_info1} \emph{local knowledge}: in each cluster, the agent $i \in \Pc_{\rm i}$  serves as the {local control center} that has the local system model of the cluster (matrices $\mb{A}^{\rm i}_{\sigt}$, $\mb{A}^{\rm ij}_{\sigt}$  and $\mb{C}_{{\rm i}_i}$) and the {local measurement} $\mb{y}_{{\rm i}_i}(t)$.
    \item \label{assum:local_info2} \emph{local measurements}: the measured output $\mb{y}_{{\rm i}_i}(t)$ in \eqref{eq:cluster_dyn2} is locally available at the node $i$ and, unlike global measurements, it is not sent to the control center to keep the output secure and inaccessible to the attacker.
    \item \label{assum:local_info3} \emph{cross-cluster communication}: every local control center, i.e., the node $i$ in cluster $\Pc_{\rm i}$, considers coupling terms $\sum_{\rm j\in \Nc_{\scriptscriptstyle {\mathcal{P}_{{\rm i}}}}} \mb{A}_{{\sigt}}^{\rm {ij}} \mb{x}_{\rm {j}}$ as unknown inputs to $\Sigma_{\scriptscriptstyle \mathcal{P}_{{\rm i}}}$. Moreover, inter-cluster couplings do not change, i.e., $\mb{A}_{{\sigt}}^{\rm {ij}}=\mb{A}_{1}^{\rm {ij}}$, $ \forall\, t \in [t_0, +\infty)$. Thus there is no need for exchange of $\mb{x}_{\rm j}$'s information between local control centers.
\end{enumerate}
\end{assumption}
The assumption \ref{assum:local_info}-\ref{assum:local_info1} is common in the literature (cf. \cite{gallo2020distributed}) as the model-based detection of cyber attacks on exchanged data over a network requires augmented knowledge of the neighboring agents' model to estimate their states and further compare them with the received data. Minimizing the local information exchange affects the scalability and depends on the sparsity of the communication network as well as on applications. 
% ==============================================
\subsection{Observer Design and Attack Detectability Analysis}
As described in Section \ref{subS:attak_detection_scheme}, the attack detection framework is composed of a centralized observer for monitoring the system \eqref{eq:cl_sys} from the control center, and a set of local observers in clusters, that serve as local attack detectors and trigger for communication topology switching. In what follows, we describe the observer design procedure based on the conditions derived in the previous section. 

\noindent
\textbf{{Decentralized observer.}} Consider the dynamics of the system partitions described in \eqref{eq:cluster_dyn2} and Assumption \ref{assum:local_info}, we use the unknown input observer (UIO) scheme in \cite{chen1996design} to estimate the cluster state $\hat{\mb{x}}_{\rm i}$ independent of the states $\mb{x}_{\rm j} $'s of the neighboring clusters (i.e. ${\rm j} \in \Nc_{\scriptscriptstyle \mathcal{P}_{{\rm i}}}$). This is achieved by considering the interconnection of local models as unknown inputs and rewriting them such that
%
% \cite{chen2006unknown} LMI for observer gain matrices
%
\begin{equation}\label{eq:E_definition}
    \sum_{{\rm j}\in \Nc_{\scriptscriptstyle \mathcal{P}_{{\rm i}}}} \mb{A}_{{\sigt}}^{\rm{ij}} \mb{x}_{\rm{j}} 
    \coloneqq
    \mb{E}_{}^{\rm i}\mb{x}^{d}_{\rm i}, \ \ {\sigt}=1, \, \forall\, t\in [t_0, +\infty),
\end{equation}
% where $\mb{E}_{}^{ i} := \col(\mb{A}_{{\sigt}}^{{j}})^T_{j \in \Nc_{\spi}}$ and $ \mb{x}^{\s d}_{i} := \col(\mb{x}_{{j}})_{j \in \Nc_{\spi}} $.
%
where $\mb{E}_{}^{\rm i}$ is a full column rank\footnote{ The columns of $\mb{E}_{}^{ \rm i}$ for cluster $\Pc_{\rm i}$ are corresponding to the edge-cuts connecting $\Pc_{\rm i}$ to its neighboring clusters.} matrix and $\mb{x}^{d}_{\rm i}$ is a vector of the states of neighboring clusters that are received by cluster $\Pc_{\rm i}$. Now, introducing the UIO state $ \mb{z}_{\rm i} =\hat{\mb{x}}_{\rm i} - \mb{h}_{}^{\rm i }\mb{y}_{{\rm i}_i} $, the dynamics of the local UIO is given by 
\begin{equation}\label{eq:obs_decent}
 \Sigma^{\s \Zc_{\rm i}}_{\s \Oc}:
 \left\{
 \begin{array}{l}
  \dot{{\mb{z}}}_{\rm i} 
  = 
  \mb{F}_{\sigt}^{\rm {i}}{\mb{z}}_{\rm i} 
%   +
%   \mb{T}_{\sigt}^{i} \mb{B}^i \mb{u}_{a_i} 
  +
  \paren{\mb{K}_{\sigt}+\bar{\mb{K}}_{\sigt}}\mb{y}_{{\rm i}_i}, 
 \\ 
 \hat{\mb{x}}_{\rm i} =\mb{z}_{\rm i}+\mb{h}_{}^{\rm i}\mb{y}_{ {\rm i}_i},\\
\hat{\mb{x}}_{\rm i}(0)=\mb{0},  \quad  \Pc_{\rm i} \subset \Vc, \quad {\rm i} \in \braces{1,\cdots,|\boldsymbol{\Pc}|},
\end{array}
\right.  
\end{equation}
where $\mb{F}_{\sigt}^{\rm {i}}, \mb{K}_{\sigt}, \bar{\mb{K}}_{\sigt}, $ and $\mb{h}_{}^{\rm i}$ are matrices 
satisfying conditions
\begin{align}\label{eq:uio_conds_1}
    \mb{T}_{}^{\rm i}&=\paren{I-\mb{h}_{}^{\rm i}\mb{C}_{{\rm i}_i}}, \ \    \paren{\mb{h}_{}^{\rm i}\mb{C}_{{\rm i}_i}-I}\mb{E}_{}^{\rm i}=0,
    \\ \label{eq:uio_conds_2}
    \mb{F}_{\sigt}^{\rm i}&=\paren{\bar{\mb{A}}_{\sigt}^{\rm {i}} -\bar{\mb{K}}_{\sigt} \mb{C}_{{\rm i}_i}}, \ \
    {\mb{K}}_{ \sigt}=\mb{F}_{\sigt}^{\rm i} \mb{h}_{}^{\rm i},  \\ \label{eq:uio_conds_3}
    \bar{\mb{A}}_{\sigt}^{\rm i} &= \mb{A}_{\sigt}^{\rm {i}} - \mb{h}_{}^{\rm i}\mb{C}_{{\rm i}_i} \mb{A}_{\sigt}^{\rm{i}}.
\end{align}
% alternative equations of UIO conditions:
%
% \begin{align}\label{eq:uio_conds_1}
% \paren{\mb{h}_{}^{ i}\mb{C}_{ n_i}-I}\mb{h}_{}^{ i}&={0},\\\label{eq:uio_conds_2}
% %
% \mb{T}_{\sigt}^{ i}&=\paren{I-\mb{h}_{}^{ i}\mb{C}_{ n_i}},\\ \label{eq:uio_conds_3}
% %
% % \mb{F}_{\sigt}^{ i}&=(\mb{A}_{\sigt}^{{i}} - \mb{h}_{}^{ i}\mb{C}_{ i} \mb{A}_{ \sigt}^{{i}} -\mb{K}_{\sigt} \mb{C}_{ i})\\
% %
% \mb{F}_{\sigt}^{ i}&=\paren{\bar{\mb{A}}_{\sigt}^{{i}} -\bar{\mb{K}}_{\sigt} \mb{C}_{ n_i}},\\ \label{eq:uio_conds_4}
% %
% {\mb{K}}_{ \sigt}&=\mb{F}_{\sigt}^{i} \mb{h}_{}^{i},
% \end{align}
%
Furthermore, $\mb{F}_{\sigt}^{\rm i}$ is Hurwitz stable over $t\in [t_0,t_{\mb{m}})$ for all \emph{normal} and \emph{safe modes}.

Consider \eqref{eq:cluster_dyn2}, \eqref{eq:obs_decent} and let $ \mb{e}_{\rm i} := \mb{x}_{\rm i} - \hat{\mb{x}}_{\rm i}$, one can use the conditions in \eqref{eq:uio_conds_1}-\eqref{eq:uio_conds_3} to obtain the error dynamics of UIO as follows
\begin{equation}\label{eq:obs_decent_error}
 \Sigma^{\mb{e}_{\rm i}}_{ \Oc}:
 \left\{
 \begin{array}{l}
 \:\dot{{\mb{e}}}_{\rm i} \:\,=  {\mb{F}_{\sigt}^{\rm {i}}} \mb{e}_{\rm i} 
      + \mb{T}_{}^{\rm i} \mb{B}^{\rm i}\mb{u}_{a_{\s \rm i}} , 
     \quad {\mb{e}}_{\rm i}(0)=\mb{x}_{\rm i}(0),
\\ 
 {\mb{r}}_{{\rm i}_i} =\mb{C}_{{\rm i}_i}\mb{e}_{\rm i} ,  \quad  \Pc_{\rm i} \subset \Vc, \ \ {\rm i} \in \braces{1,\cdots,|\boldsymbol{\Pc}|}.
\end{array}
\right.  
\end{equation}
%
% \begin{equation}\label{eq:obs_decent_error}
%  \Sigma^{\mb{e}_{ i}}_{ \Oc}:
%  \left\{
%  \begin{array}{l}
%  \dot{{\mb{e}}}_{i} =  {\mb{F}_{\sigt}^{{i}}} \mb{e}_{i} 
%       + \mb{T}_{\sigt}^{i} \mb{B}^i\mb{u}_{a_{\s i}} 
%      + \bar{\mb{K}}_{\sigt} \bar{\mb{u}}_{s_{\s i}} + \mb{h}_{}^i \dot{\bar{\mb{u}}}_{s_{\s i}} , 
% \\ 
%  {\mb{r}}_i =\mb{C}_{i}\mb{e}_i -\bar{\mb{u}}_{s_{ i}} ,\\
%  {\mb{e}}_{ i}(0)=\mb{x}_{ i}(0),  \quad  \Pc_i \subset \Vc, \quad i \in \braces{1,\cdots,|\boldsymbol{\Pc}|}.
% \end{array}
% \right.  
% \end{equation}
%
In the absence of adversarial attacks, $\mb{u}_{a_{\s \rm i}}=\boldsymbol{0}$, it is straightforward to show that  $ \lim_{t\rightarrow \infty} \mb{e}_{\rm i}(t) = \boldsymbol{0}$ as $\mb{F}_{\sigt}^{\rm i}$ is Hurwitz stable in all modes. LMI-based approaches can be used to design \eqref{eq:uio_conds_2} such that \eqref{eq:obs_decent_error} remains stable under arbitrary switching \cite{chesi2011nonconservative}. 

Recall Assumption \ref{assum:local_info}-\ref{assum:local_info2}, unlike the case of global measurements (cf. Proposition \ref{propo:stealthy_attacks}-({\rm{i}})), the local measurements $\mb{y}_{{\rm i}_i}$'s are hidden and thus cannot be altered by the attacker to cancel out the effect of the attack $\mb{u}_{a_{\rm i}}$ on the output of \eqref{eq:cluster_dyn2}. This difference also manifests itself in the residual of local observer \eqref{eq:obs_decent_error}. Therefore, in order to determine the stealthiness of attack $\mb{u}_{a_{\rm i}}$ with respect to the local residual signal $\mb{r}_{{\rm i}_i}$, it is necessary and sufficient to investigate whether the stealthiness conditions presented in Proposition \ref{propo:stealthy_attacks} are satisfied for the system in \eqref{eq:obs_decent_error}.

%\Hamid{The last sentence in this paragraph is a little unclear!}. \Mo{modified}

In the following proposition, we formally characterize the conditions for the detection of stealthy attacks using the local observer in \eqref{eq:obs_decent}. 

\begin{proposition}\btitle{Attack detectability of local observers}\label{prop:attack_detectability_local}
For a strongly connected cluster $\Pc_{\rm i}$ with $\boldsymbol{\Ec}$ inter-clustering edges and $|{\overline{\Fc_{\rm i}}}|$ compromised agents, there exists a local observer given by \eqref{eq:obs_decent} to locally detect the stealthy attacks if 
\begin{enumerate}
    \item  there is a $\mb{k}$-connected node $i \in \Pc_{\rm i}$ as the local monitored agent such that $\mb{k} \geq \boldsymbol{\Ec} + |{\overline{\Fc_{\rm i}}}|$,
    \item  $ \rank\paren{\mb{C}_{{\rm i}_i}\mb{E}_{}^{\rm i}}=\rank\paren{\mb{E}_{}^{\rm i}}$,
    \item  the matrix pencil $\mb{P}$ in \eqref{eq:pencil_uio} is full (column) rank,
\begin{align}\label{eq:pencil_uio}
\mb{P} =
    \begin{bmatrix}
        \lambda_0 I_{} -{\mb{A}}^{\rm i}_{\sigt} &  -{\mb{B}}^{\rm i} & {\mb{E}}^{\rm i}_{}
        \\
        \mb{C}_{{\rm i}_i} & 0 & 0
    \end{bmatrix}.
\end{align}
\end{enumerate}
where the tuple $\paren{{\mb{A}}^{\rm i}_{\sigt},{\mb{B}}^{\rm i},\mb{C}_{{\rm i}_i}}$ and matrix ${\mb{E}}^{\rm i}_{}$  are defined in \eqref{eq:cluster_dyn2} and \eqref{eq:E_definition}, respectively. 
\end{proposition}
\begin{IEEEproof}
See Appendix \ref{app_attack_detectability_local}.
\end{IEEEproof}

\begin{remark}\btitle{Evaluation of the condition in \eqref{eq:pencil_uio}}\label{rmk:local_obs_measurements}
% \btitle{Practicability of Proposition \ref{prop:attack_detectability_local}} 
Conditions \emph{(i)}-\emph{(iii)} in Proposition \ref{prop:attack_detectability_local} are equivalent to necessary and sufficient conditions for the existence of UIO in \eqref{eq:obs_decent} \cite{chen1996design}. 
% Moreover, condition \emph{(i)} determines the maximum tolerable number of compromised actuator channels, as a detection bound, for any given $\mb{k}$-connected node $n_i$. 
It is worth noting that as matrix $\mb{B}^{\rm i}$ in \eqref{eq:pencil_uio} is unknown to the defender, it can be replaced with $I_{N_i}$, i.e., assuming all the nodes of the cluster are under attack, in analysis and selecting locally monitored agents associated with $\mb{C}_{{\rm i}_i}$. This, however, may require further communication between agents within a cluster. Alternatively, as in a set cover problem setting, a set of local monitoring agents that each of them satisfies the conditions \emph{(i)}-\emph{(iii)} for part of a cluster can be used to cover all of nodes of the cluster \cite{teixeira2014distributed}. Minimizing the number of local measurements versus the number of local observers is a trade-off problem which will be the subject of future work.
\end{remark}
% A set of local monitoring agents satisfying the conditions in  \ref{*} can be obtained by solving the following set cover problem
% % \begin{subequations}
% \begin{align}
%  &   \min_{\Mc_i \subseteq \Pc_i} &&   |\Mc_i| \nonumber\\ 
%  & \text{subject to} && \bigcup_{j \in \Mc_i} \Nc_j = \Pc_i, \quad i \in \braces{1,\cdots,{|\boldsymbol{\Pc}|}}.
% \end{align}
% \end{subequations}
%
% \begin{remark}[mode observability and state observability]
% \end{remark}
%%%%%%%%%%%%%%%%%%%%%%%%%%%%%%%%%%%%%%%%%%%%%%%%%%%%%%%%%%%%%%%%%%%%%%%%%%%%%%%%

\noindent
\textbf{{Centralized observer.}} Consider the dynamical system \eqref{eq:cl_sys}, a Luenberger-type centralized observer, derived based on the normal mode $\sigt=1$,  is given by
\begin{equation}\label{eq:obs_cent}
\Sigma_{\s \Oc}^{\s \Mc} :
\left\{
\begin{array}{ll}
\dot{\hat{\mb x}} 
=
\mb{A}_{\sigt} \hat{\mb x} + \mb{H}_{\sigt}(\mb{y}_{\s} - \hat{\mb y}_{} ),
& \sigt=1, \\
\hat{\mb{y}}_{\s }=\mb{C}_{\s }\hat{\mb x}, 
% 
% & \hat{\mb{x}}(0)=\hat{\mb{x}}_0, \\ 
& \hat{\mb{x}}(0)=\boldsymbol{0}, \\
\mb{r}_{\s 0} = (\mb{y}_{\s } -\hat{\mb{y}}_{\s }), & \text{residual,}
\end{array}
\right.  
\end{equation}
where $ \mb{H}_{\sigt}$ is the observer gain and $\mb{r}_{\s 0}(t)$ denotes the residual signal available in the control center for monitoring purposes. 

In order to design the observer gain $ \mb{H}_{\sigt}$, the partial observability of pair $(\mb{A}_{ \sigt},\mb{C})$ imposed in Section \ref{Ss:privacy} and the activated mode $\sigt$ should be taken into account. An immediate solution is to define an LMI optimization problem finding a constant $\mb{H}_{\sigt} \coloneqq \mb{H}$ by which $(\mb{A}_{\sigt}-\mb{H}\mb{C})$ is (Hurwitz) stable in all modes \cite{chilali1996h,chen2004observer}. 
% In \cite{mao2020novel} a special solution was used for the case that the initial conditions are available (i.e. $\hat{\mb{x}}(0)={\mb{x}}(0)$).

From Assumption \ref{assum:attacker} and condition \eqref{eq:stealthy}, it is straightforward to show that 
the attack $\mb{u}_a$ remains stealthy for the observer \eqref{eq:obs_cent} in the normal mode over the time span $t\in[t_0,t_1)$ where $\mb{A}_{\sigt}=\mb{A}_{1}$.

Recall \eqref{eq:covert_trajec} and \eqref{eq:zda_trajec}, and let
\begin{align}\label{eq:est_error_normal}
    \bar{\mb{e}} 
    &\coloneqq 
    \bar{\mb{x}}-\hat{\mb{x}} 
    \\ \label{eq:est_error}
    \mb{e} 
    &\coloneqq 
    \mb{x}-\hat{\mb{x}}=\bar{\mb{x}}+\tilde{\mb{x}}-\hat{\mb{x}}=\bar{\mb{e}}+\tilde{\mb{x}}
\end{align}
be the estimation error of the states of an attack-free system ($\dot{\bar{\mb{x}}}=\mb{A}_{\sigt}\bar{\mb{x}},\, \mb{y}=\mb{C}\bar{\mb{x}}$) and the under attack system in \eqref{eq:cl_sys}, respectively.
Then using \eqref{eq:cl_sys} and \eqref{eq:obs_cent}, the error dynamics of the centralized observer is given by 
\begin{equation}\label{eq:obs_cent_error}
\Sigma_{\s \Oc}^{\mb{e}} :
\left\{
\begin{array}{l}
\dot{{\mb e}} 
=
(\mb{A}_{1}-\mb{H}\mb{C}) {\mb e} 
+ (\mb{A}_{\sigt}-\mb{A}_{1}){\mb{x}}
+\mb{H}\mb{u}_s+ \mb{B}\mb{u}_a, \\ 
{\mb{e}}(0) = \mb{x}_0,\\ % ={\mb{e}}_0
\mb{r}_{\s 0} = (\mb{y}_{\s } -\hat{\mb{y}}_{\s })=\mb{C}\mb{e}-\mb{u}_s=\mb{C}\bar{\mb{e}}, \quad \text{ residual},
\end{array}
\right.  
\end{equation}
where for measurement $\mb{y}$ in \eqref{eq:obs_cent} we used the expression $\mb{y}=\mb{C}\mb{x}-\mb{u}_s$ as defined in \eqref{eq:cl_sys}. Consider \eqref{eq:attack_model} and \eqref{eq:stealthy}, $\mb{y}$ in \eqref{eq:obs_cent} also satisfies $\mb{y}=\mb{C}\mb{x}-\mb{u}_s=\mb{C}\mb{x}-\mb{C}\tilde{\mb{x}}=\mb{C}\bar{\mb{x}}$. Then using $\mb{y}=\mb{C}\bar{\mb{x}}$, \eqref{eq:cl_sys}, \eqref{eq:attack_model}, \eqref{eq:obs_cent}, \eqref{eq:est_error_normal}, the following dynamics is obtained
\begin{equation}\label{eq:obs_cent_error_normal}
\Sigma_{\s \Oc}^{\bar{\mb{e}}} :
\left\{
\begin{array}{l}
\dot{\bar{\mb e}} 
=
(\mb{A}_{1}-\mb{H}\mb{C}) \bar{\mb e} 
+ (\mb{A}_{\sigt}-\mb{A}_{1})\bar{\mb{x}}, \\ 
\bar{\mb{e}}(0) = \bar{\mb{x}}_0,\\ % ={\mb{e}}_0
\bar{\mb{r}}_{\s 0} =\mb{C}\bar{\mb{e}}, \quad \text{ residual}.
\end{array}
\right.  
\end{equation}
Note that, during normal mode $\sigt=1$ over the time span $\forall \, t \in [t_0, t_1)$, the residual $\mb{r}_{\s 0}$ in \eqref{eq:obs_cent_error} is the same as that of \eqref{eq:obs_cent_error_normal} that is the dynamics of the estimation error of system states in the absence of attacks. This implies that, in the case of a covert attack with $\mb{u}_s \neq 0 $, as long as signal $\mb{u}_s(t)$ cancels out the effect of $\mb{u}_a(t)$ on the output $\mb{y}(t)$, the residual $ \mb{r}_0(t) = \mb{C}\bar{\mb{e}}(t)$ converges to zero as $t_1 \rightarrow +\infty$, yielding the stealthiness of the covert attack, in the normal mode, for the centralized observer \eqref{eq:obs_cent}. 
\\
In the case of a ZDA, $\mb{u}_s = 0$ in \eqref{eq:obs_cent_error} although \eqref{eq:stealthy} still holds that leads to the stealthiness of a ZDA for the observer \eqref{eq:obs_cent}. To show this, one need to verify the attack $\mb{u}_a$ remains in the zeroing direction of \eqref{eq:obs_cent_error}. Using Definition \ref{def:inv_zeros} for \eqref{eq:obs_cent_error} in the normal mode, we obtain 
\begin{equation}\label{eq:obs_cent_INV_zero}
    \begin{bmatrix}
        \lambda_0 I_{} - (\mb{A}_{1}-\mb{H}\mb{C}) & -\mb{B} \\ \mb{C} & 0_{}
    \end{bmatrix}
    \begin{bmatrix}
        \tilde{\mb{e}}(0) \\ \mb{u}_0
    \end{bmatrix}
    =
    \begin{bmatrix}
        \boldsymbol{0} \\ \boldsymbol{0}
    \end{bmatrix},
\end{equation}
where $\tilde{\mb{e}}(0) \coloneqq \mb{e}(0)-\bar{\mb{e}}(0)=\mb{x}_0-\bar{\mb{x}}_0= \tilde{\mb{x}}_0$. Recall $\tilde{\mb{x}}_0 \in \ker(\mb{C})$ in \eqref{eq:zda_cond}, then the second equation of \eqref{eq:obs_cent_INV_zero} yields $\mb{C}\tilde{\mb{e}}(0)=\mb{C}\tilde{\mb{x}}_0=\boldsymbol{0}$. Applying $\mb{C}\tilde{\mb{e}}(0)=\boldsymbol{0}$ into the first equation of \eqref{eq:obs_cent_INV_zero} simplifies the matrix pencil in \eqref{eq:obs_cent_INV_zero} into that of \eqref{eq:zda_cond} over $t\in[t_0,t_1)$ where $\mb{A}_{\sigt}=\mb{A}_{1}$. This ensures the stealthiness of ZDA in the normal mode for the observer \eqref{eq:obs_cent}.

The following Theorem provides conditions to address Problem \ref{prob:attack_detection}-(ii) by characterization of switching modes that lead to attack detection with respect to global measurements.
\begin{theorem}\btitle{Attack detectability under switching communication}\label{th:Attack_detectability_switching}
Consider system \eqref{eq:cl_sys} under stealthy attack modeled in \eqref{eq:attack_model}, and let intra-cluster topology switching satisfy
% to a safe attack detection if it satisfies
%
\begin{enumerate} 
    \item $ \Img (\Delta \Lc_{\mb{q}}) \cap  \ker{([\mb{C}^{\top}_{\rm x}\; \mb{C}^{\top}_{\rm v}]^{\top})} =\emptyset,$
    % \item $\exists \, \Delta \Lc_{\mb{q}} \ \  \rm{s.t.} \ \ \Img (\Delta \Lc_{\mb{q}}) \cap  \ker{([\mb{C}^{\top}_{\rm x}\; \mb{C}^{\top}_{\rm v}]^{\top})} =\emptyset,$
    % \label{cond:detection_condition_necc}
    %
    % \item $\Dc_{\rm i} \cap \Mc \neq \emptyset, \, \forall \, {\rm i} \in \braces{1,\cdots,|\boldsymbol{\Pc}|}$,
    \label{cond:detection_condition_suff_0}
    \item $\Lc_{\mb{q}}$ features distinct eigenvalues,
    \label{cond:detection_condition_suff_1}
    \item $[\Uc_{\mb{q}}]_{i,\ell}-[\Uc_{\mb{q}}]_{j,\ell} \neq 0,\ \ \forall \, \ell \in \Vc\setminus\{1\}, \, \forall \, i,j \in \Dc_{\rm c},\, \forall \, {\rm c} \in \{1,\cdots, \boldsymbol{{\rm c}}\} $,\label{cond:detection_condition_suff_2}
\end{enumerate}
where $\Delta \Lc_{\mb{q}} \! \coloneqq \! \Lc_{\sigt}-\Lc_{1} $, with $ \sigt \!=\! \mb{q} \! \in \!  \Qc,$ $ t \in [t_1, +\infty)$, $\mb{C}^{\top}_{\rm x}$ and $ \mb{C}^{\top}_{\rm v}$ are given in \eqref{eq:cl_sys}-\eqref{eq:measurments} and $\Dc_{\rm c} \!\subset \! \Vc,$ denotes the set of nodes in ${\rm c}$-th connected component of $\Delta \Lc_{\mb{q}}$ corresponding to agents involved in connected switching links, and finally $\Uc_{\mb{q}}$ is a unitary matrix ( $\!\Uc_{\mb{q}}\Uc_{\mb{q}}^{\top}=I$) diagonalizing Laplacian $\Lc_{\mb{q}}$.
\\
% Then during achieving the consensus task in \eqref{eq:cond_consensus}, topology switching leads to attack detection through centralized observer \eqref{eq:obs_cent}. 
Then, ZDA and covert attacks undetectable for the centralized observer \eqref{eq:obs_cent} are impossible only if the topology switching satisfies conditions {\ref{cond:detection_condition_suff_0}-\ref{cond:detection_condition_suff_2}}. If additionally the system is not at its exact consensus equilibrium when the attack is launched,  conditions {\ref{cond:detection_condition_suff_0}-\ref{cond:detection_condition_suff_2}} are sufficient for the detection of ZDA.  
\end{theorem}
\begin{IEEEproof} See Appendix \ref{app_Attack_detectability_switching}. \end{IEEEproof}
\begin{remark}\btitle{Safe topology switching}\label{rmk:Safe_topology_switching}
% \btitle{Practicability of conditions in Theorem \ref{th:Attack_detectability_switching}} 
For a given pair $(\mb{A}_{\sigt},\mb{C})$ in \eqref{eq:cl_sys}, one can compute a set of switching modes by evaluating the conditions {\ref{cond:detection_condition_suff_0}-\ref{cond:detection_condition_suff_2}} of Theorem \ref{th:Attack_detectability_switching}. This could be performed through iterative algorithms changing graph connections. 
%
% Moreover, although identifying $\boldsymbol{\Zc}$ is practically impossible as $\mb{B}$ and $\tilde{\mb{x}}_0$ in \eqref{eq:attack_model} are unknown to the defender, local observers detecting stealthy attacks in a cluster can locally identify and trigger a safe switching mode that satisfies {\ref{eq:unsafe_set}}.
%
Furthermore, if $ \boldsymbol{\Zc} $ be an unknown subspace associated with system states affected by stealthy attack $\mb{u}_a(t)$ i.e. $\tilde{\mb{x}}(t) \in \boldsymbol{\Zc}$. Then, in view of $\tilde{\mb{x}}_0 = \mb{x}_0- \bar{\mb{x}}_0$ (see Proposition \ref{propo:stealthy_attacks}), the discrepancy term $(\mb{A}_{\sigt}-\mb{A}_{1}){\mb{x}}$ in the dynamical system \eqref{eq:obs_cent_error} will be bounded and vanishing if
\begin{equation}\label{eq:unsafe_set}
 \boldsymbol{\Xc}_{\mb{q}} \cap  \boldsymbol{\Zc} =\emptyset.
\end{equation}
Therefore, if condition \eqref{eq:unsafe_set} holds,  $(\mb{A}_{\sigt}-\mb{A}_{1}){\mb{x}}$ does not effect the stability of the system, as a consequence of input-to-state stability property of consensus systems \cite{meng2018consensus}.
% This result is consistent with Problem \ref{prob:attack_detection}-\emph{(ii)}. 
It is also noteworthy that although identifying $\boldsymbol{\Zc}$ beforehand is practically impossible as $\mb{B}$ and $\tilde{\mb{x}}_0$ in \eqref{eq:attack_model} are unknown to the defender, local observers detecting stealthy attacks in a cluster can locally identify and trigger a safe switching mode that satisfies \eqref{eq:unsafe_set}.
\end{remark}
%
% \begin{remark}\btitle{On attack identification} 
% Condition {\ref{eq:detection_condition}} in Theorem \ref{th:Attack_detectability_switching} specifies column spaces that any change in their respective direction manifests itself in system outputs. Therefore, it is possible to inject distinguishable signal though the direction as signatures showing the exact location of compromised agents in a cluster. Also it allows to design safe mode switches such that topology switches in clusters only are detectable only in  globally monitored agents the respective clusters. {\ref{eq:unsafe_set}}. 
% \end{remark}
%
\subsection{Attack Detection Procedure}\label{S:detection}
%
% \Mo{maybe we can put Proposition III.2 and Theorem III.3 into this section can rename the Section to Attack detection analysis??}\Hamid{You can divide section III into two parts: III. Attack detection Framework (including parts A-C) and Section VI. Observer Design (part D). }

The results in the previous section provide conditions for the detectability of stealthy attacks locally, at the cluster level, and globally, at a ground control station equipped with a centralized observer. As described earlier, the attack detection framework relies on switching communication links  generating a discrepancy between the attacker model \eqref{eq:attack_model} and the actual system \eqref{eq:cl_sys}. To this end, at local level (clusters), unknown-input observers in \eqref{eq:obs_decent}, satisfying conditions of Proposition \ref{prop:attack_detectability_local}, locally detect stealthy attacks. Followed by the detection, a local observer triggers a topology switching, $\Gc_{\sigt}$, that satisfies conditions \ref{cond:detection_condition_suff_0}-\ref{cond:detection_condition_suff_2} of Theorem \ref{th:Attack_detectability_switching}, yielding stealthy attack detection in the control center. This procedure is depicted in Algorithm \ref{alg:detection_procedure}.
\begin{algorithm}[H]
\caption{Topology switching for attack detection}\label{alg:detection_procedure}
\begin{algorithmic}[1]
\Procedure{Attack detection}{$\,\Gc_{\sigt}$, Obs. in \eqref{eq:obs_cent}, \eqref{eq:obs_decent}}       %\Comment{This is a test}
  	\State \textbf{do} run global observer \eqref{eq:obs_cent} and local observers \eqref{eq:obs_decent}.
    \If{$\mb{r}_{{\rm i}_i}(t) > \texttt{threshold} $}
        \State \textbf{do}  Identify a safe mode $\sigt=\mb{q}\in \Qc$ for $\Lc_{\sigt}$ that satisfies conditions \ref{cond:detection_condition_suff_0}-\ref{cond:detection_condition_suff_2} in Theorem. \ref{th:Attack_detectability_switching}
        \State \textbf{do} Trigger an identified safe mode $\sigt=\mb{q}\in \Qc$
    \If{$ \mb{r}_0(t) > \texttt{threshold} $}
        \State Stealthy attack is detected.
        % \ElsIf{$Condition \neq 5$}
    \EndIf
    % \Else
    % \State System operating in the normal mode.
    \EndIf
\EndProcedure
\end{algorithmic}
\end{algorithm}
As presented in Algorithm \ref{alg:detection_procedure}, the observers (attack detectors) require an appropriate threshold for their residuals to avoid false attack detection. These thresholds can be designed by considering an upper bound on the estimation error of observers in the attack-free case. An analytical analysis, however, will be the subject of future work.

% \begin{remark}\btitle{Comparison with literature}\label{rmk:comparision}
% In terms of scalability, our framework does not require full knowledge of system model $(\mb{A}_{\sigt},\mb{C})$ in each node for attack detection (cf. \cite{sundaram2010distributed, teixeira2010networked, teixeira2014distributed}). In terms of the generality of detection framework, it is not required for communication topology of system \eqref{eq:cl_sys} to have a Laplacian matrix with distinct eigenvalues (cf. \cite{mao2020novel})
% \end{remark}

%%%%%%%%%%%%%%%%%%%%%%%%%%%%%%%%%%%%%%%%%%%%%%%%%%%%%%%%%%%%%%%%%%%%%%%%%%%%%%%%
\section{Simulation Results}\label{S:Results}
\begin{figure*}[t]
  \centering \subfloat[Case 1: bounded residual]{\small
    \hspace*{-10pt} \includegraphics[width=.25\linewidth]{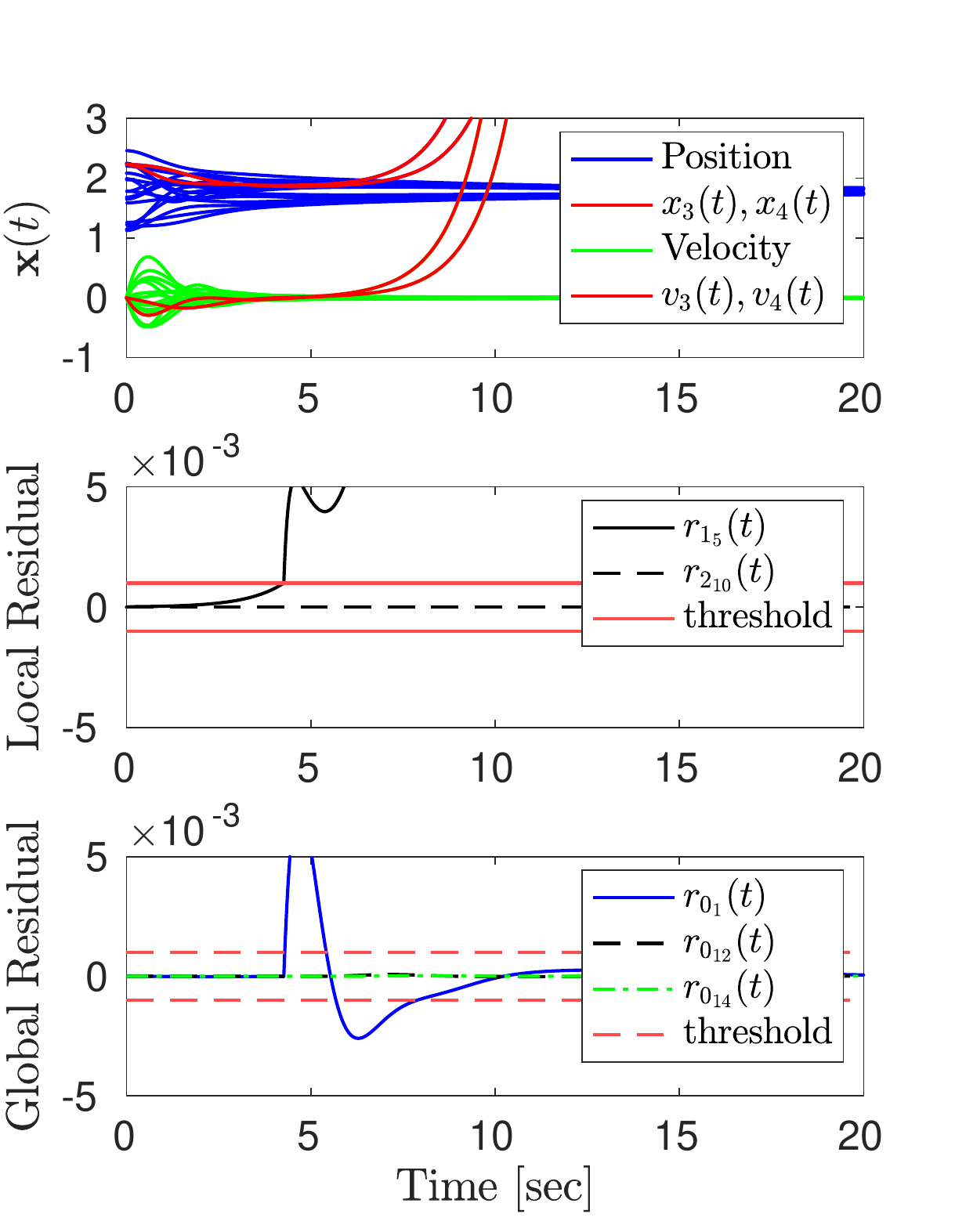}
  }
  \subfloat[Case 2: unbounded residual]{\small
    \hspace*{-10pt} \includegraphics[width=0.25\linewidth]{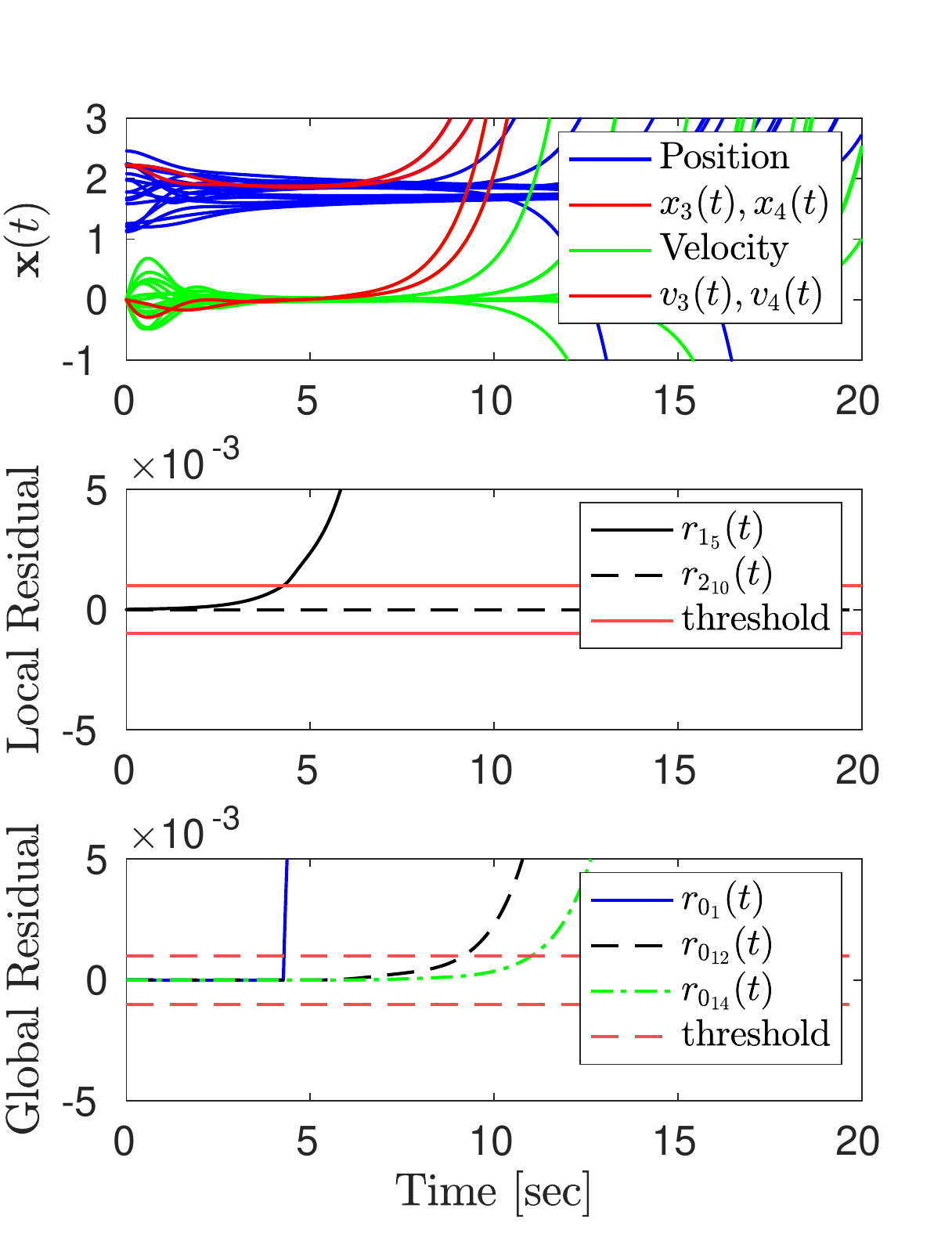}
  }
  \subfloat[Case 3: no detection]{\small
    \hspace*{-10pt} \includegraphics[width=0.25\linewidth]{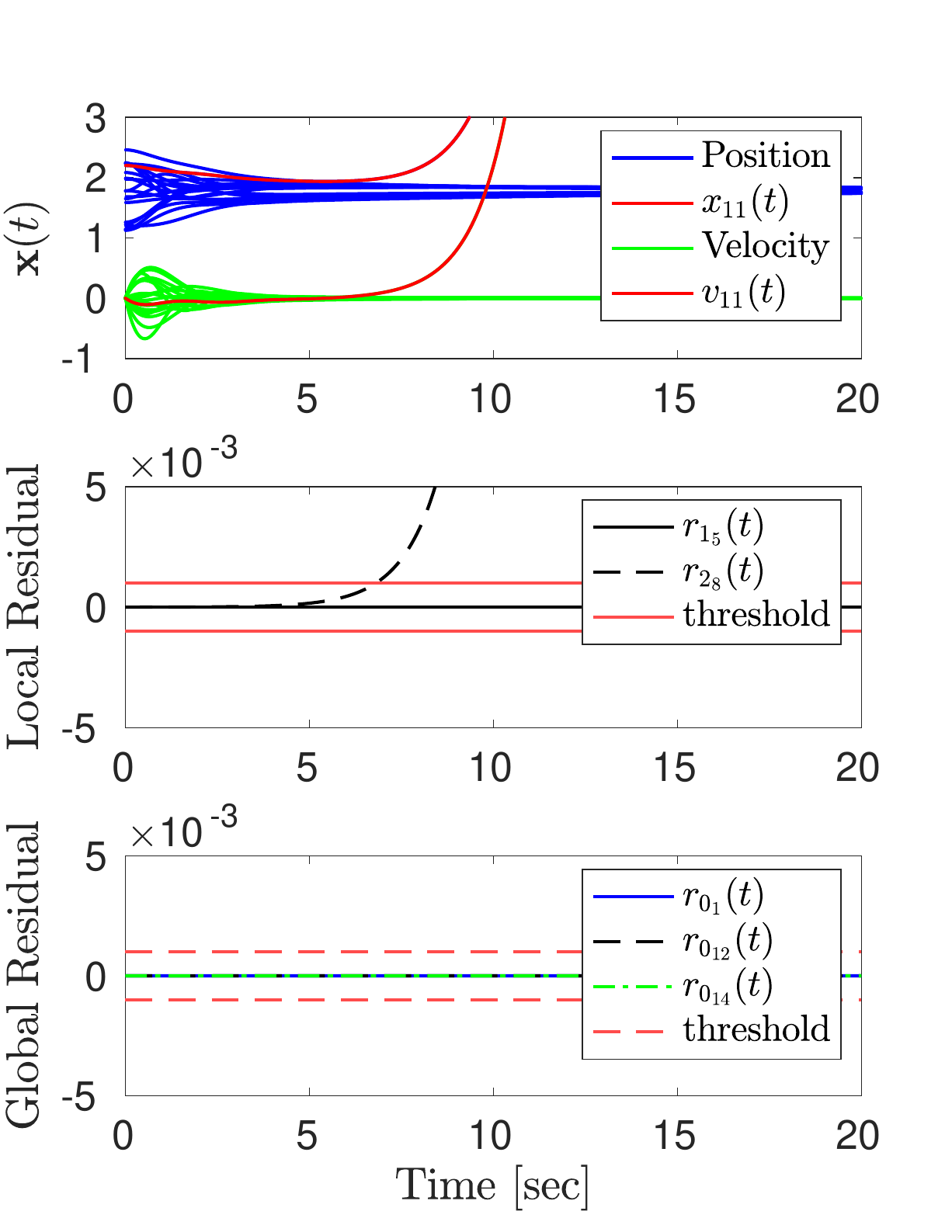}
  }
  \subfloat[Communication topology]{\small 
  \includegraphics[width=0.25\linewidth]{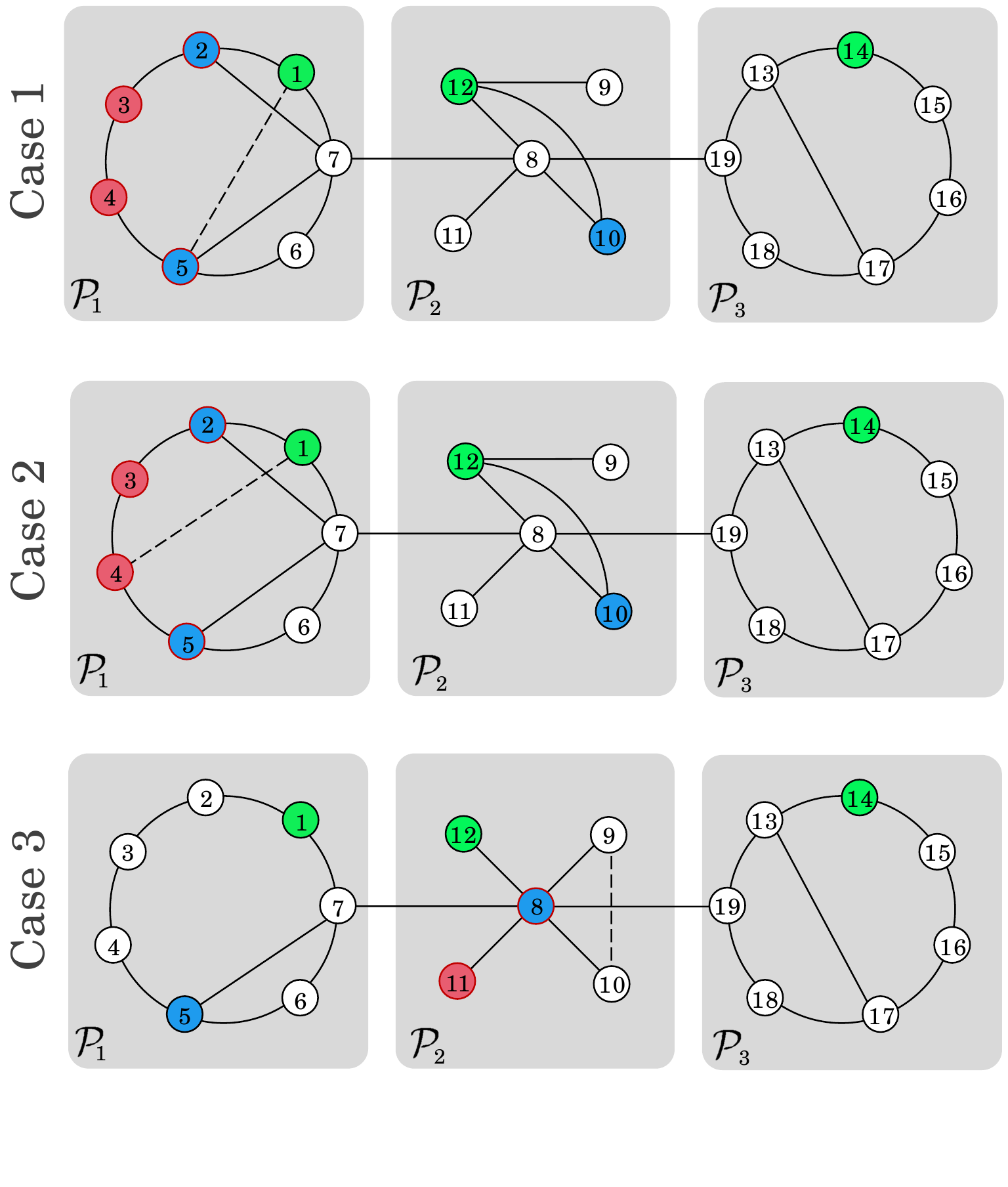}}
%   \subfloat[Case 4: no detection]{\small
%   \begin{tabular}[b]{c@{}} 
%   \subfloat{\small 
%   \includegraphics[scale=0.28]{figures/C1.pdf}}\\
%   %
%   \subfloat{\small
%   \includegraphics[scale=0.28]{figures/C2.pdf}}\\ 
%   %
%   \subfloat[Communication topology \label{subfig:topo}]{\small 
%   \includegraphics[scale=0.28]{figures/C3.pdf}} 
%  \end{tabular}
    \hspace*{-10pt} 
  %
%   \\
% \subfloat[Case 1]{\small
%     % \hspace*{-5pt}
%     \includegraphics[width=0.23\linewidth]{figures/G1.png}
%   }
% \subfloat[Case 2]{\small
%     \hspace*{2pt}
%     \includegraphics[width=0.23\linewidth]{figures/G2.png}
%   }
% \subfloat[Case 3]{\small
%     \hspace*{2pt}
%     \includegraphics[width=0.23\linewidth]{figures/C1.pdf}
%   } 
% \subfloat[Case 4]{\small
%     \hspace*{2pt}
%     \includegraphics[width=0.23\linewidth]{figures/G4.png}\label{f:nn}
%   }  
  \caption{\small Simulation results of privacy-preserving stealthy attack detection for a $19$-node multi-agent control system. The state trajectory $\mb{x}(t)$ consists of the agents' position (blue) and velocity (green) as well as red trajectories showing affected agents by the stealthy attack (ZDA). (a)-(c) the results of attack detection for three cases with their respective control topology switching depicted in (d). In all cases of (d), green nodes show the globally monitored agents by the centralized observer, blue nodes indicate the local control centers equipped with local observers, and red-bordered shows compromised agents and red-coloured nodes indicate compromised agents affected by the stealthy attack (ZDA). Finally, dashed lines (edges) show the switching communication links. In the local residual's figures, with slight abuse of notations (cf. \eqref{eq:obs_decent_error}), the scalar residual $\mb{r}_{{\rm i}_i}$ shows only the velocity estimation error of node $i$.} 
  \label{fig:simulation_results}
\vspace*{-1ex}
\end{figure*}
We use a numerical example to validate the performance of the attack detection framework. We consider a network of $N=19$ agents and investigate, in three cases, the effect conditions proposed in Proposition \ref{prop:attack_detectability_local} and Theorem \ref{th:Attack_detectability_switching} on stealthy attack detection. It is assumed that the network has been partitioned into three clusters $\Pc_1=\braces{1,\cdots,7}$, $\Pc_2=\braces{8,\cdots,12}$, $\Pc_3=\braces{13,\cdots,19}$. Each cluster is equipped with the local observer \eqref{eq:obs_decent} (specified by blue nodes in Figure \ref{fig:simulation_results}) whose local measurements are consistent with Assumption \ref{assum:local_info} and Proposition \ref{prop:attack_detectability_local}. More specifically, In cases 1 and 2, cluster $\Pc_1$ has two local observers that each has access to its neighboring agents' measurements. In cluster $\Pc_2$, however, we considered one local observer having more communication with other agents within the cluster for its realization (cf. Remark \ref{rmk:local_obs_measurements}). Similar analysis is applied to case 3. Moreover, there is a centralized observer with global measurements as $\Mc_{x}=\emptyset$, $\Mc_{v}=\braces{7,12,14}$ consistent with Lemma \ref{lemma:Inv_unobs_subspace}. In the simulations, the system's initial conditions are considered to be known for observers although this is not a requirement for the presented theoretical results. Also, the constant thresholds were selected by evaluating the observers' performance in different case studies.

In cases 1 and 2 (shown respectively in Figures \ref{fig:simulation_results}-(a) and \ref{fig:simulation_results}-(b) with their communication topology in Figure \ref{fig:simulation_results}-(d)) a ZDA occurs in cluster $\Pc_1$ and particularly affects agents $3$ and $4$. As depicted, ZDA is stealthy in the global residuals $\mb{r}_{0_i}\text{'s}, \, i\in\braces{1,12,14}$ before topology switching. It is, however, detectable in local residual $\mb{r}_{1_5}(t)$. The local control center, node $5$, can trigger either of case 1's or case 2's switching topologies shown in Figures \ref{fig:simulation_results}-(d). While the conditions \ref{cond:detection_condition_suff_0}-\ref{cond:detection_condition_suff_2} of Theorem \ref{th:Attack_detectability_switching} are met in both cases, only case 2 meets \eqref{eq:unsafe_set} of remark \ref{rmk:Safe_topology_switching}. Consequently, the global residual $\mb{r}_{0_1}(t)$ for case 1 is bounded and vanishing after topology switching while that of case 2 is unbounded.

In cases 3 (shown in Figure \ref{fig:simulation_results}(c) with its communication topology in \ref{fig:simulation_results}-(d)) a ZDA occurs in cluster $\Pc_2$ and particularly affects agents $11$. Note that, unlike in cases 1 and 2, none of the Theorem \ref{th:Attack_detectability_switching}'s conditions are met in case 3, yielding the global residuals $\mb{r}_{0_i}(t), \, i\in\braces{1,12,14}$ remain unaffected
by the switching topology. Consequently, stealthy attack is not detectable.

Moreover, comparing cases 1's bounded global residual with case 2's unbounded global residual, it is noteworthy that meeting condition \eqref{eq:unsafe_set} yields a trade-off between a faster attack detection at a price of further exposing system states to ZDA and a slower detection by keeping uncompromised system states bounded.
\section{Conclusions}\label{S:conclusion}
In this paper, a novel attack detection framework is developed to detect   stealthy attacks against a class of multi-agent control systems seeking average consensus. The scalability of the approach is addressed using decentralized local observers. Also, the privacy preservation of the multi-agent system's state information is achieved by imposing unobservability conditions for the central (global) observer. Theoretical conditions were derived for the detectability of the stealthy attacks. The numerical example validates the theoretical results and illustrates the effectiveness of the proposed approach. 
Also, a discussion was provided on different types of switching topologies and their outcome for stealthy attack detection. 
% that constitutes to a safe or detrimental switching. 
% Formal verification of safe switching topology
Deriving sufficient and verifiable conditions on safe topology switching
as well as optimizing the number of local observers and their respective measurements will be subjects of future work.


%%%%%%%%%%%%%%%%%%%%%%%%%%%%%%%%%%%%%%%%%%%%%%%%%%%%%%%%%%%%%%%%%%%%%%%%%%%%%%%%

%%%%%%%%%%%%%%%%%%%%%%%%%%%%%%%%%%%%%%%%%%%%%%%%%%%%%%%%%%%%%%%%%%%%%%%%%%%%%%%%

%%%%%%%%%%%%%%%%%%%%%%%%%%%%%%%%%%%%%%%%%%%%%%%%%%%%%%%%%%%%%%%%%%%%%%%%%%%%%%%%
% \section*{Appendix}
\appendices
% \section{Additional Definitions and Lemmas}\label{appdx:Additional_Lemmas}
\section{}\label{appdx:Additional_Lemmas}
The followings are used in the Proof of Theorem \ref{th:Attack_detectability_switching}.
\begin{definition}\label{prop:app_def_diagonal}
The Laplcaian matrix of the graph composed of switching links between two communication graphs is block diagonalizable,
% Also each block encodes the connected (added/removed) switching links.
where each block, also called a component, encodes either a single (added/removed) switching link or a group of them are are connected.
%
% let $\Delta \Lc_{\mb{q}} = \Lc_{\mb{q}} - \Lc_{1}$ denote the difference of the Laplacian matrices in safe and normal mode. Then
% %
%      $\Delta \Lc_{\mb{q}}$ can be represented in block-diagonal form as
%     %
%     \begin{align}
%     \hspace{-1ex}
%     \breve{\Lc}_\mb{q} 
%     =
%      \diag{ \{\Delta \Lc_{\mb{q}}(\Dc_{\rm 1}), \cdots, \Delta \Lc_{\mb{q}}(\Dc_{\boldsymbol{\rm c}}), \Delta \Lc_{\mb{q}}(\Dc_{\rm s}) \}}, 
%     \end{align}
%     %
%     where $\!\Delta \Lc_{\mb{q}}(\Dc_{\rm c})\!$ denotes the Laplacian matrix of the ${\rm c}$-th connected graph component among $\!\boldsymbol{\rm c}\!$ connected components (cf. Definition \ref{def:graph_component}) with $\Dc_{\rm c}\!$ denoting the set of nodes (agents involved in connected switching links) corresponding to ${\rm c}$-th connected component, and $\!\Delta \Lc_{\mb{q}}(\Dc_{\rm s}) \!=\! 0$ denotes the Laplacian matrix of the singletons with no connections that are grouped in set $\!\Dc_{\rm s}$.
%     %
\end{definition}
%
% \begin{IEEEproof}
%
The above definition can be formally presented as follows: consider a network topology switching between two graphs with Laplacian matrices $\Lc_{\sigt=\mb{q}'}$ and  $\Lc_{\sigt=\mb{q}},\, \mb{q}', \mb{q} \in  \Qc,\, \mb{q}' \neq \mb{q}$ and let $\Delta \Lc_{\mb{q}} = \Lc_{\mb{q}} - \Lc_{\mb{q}'}$ denote the difference of their Laplacian matrices. Then, under Definition \ref{def:graph_component},
$\Delta \Lc_{\mb{q}}$ is associated with the induced graph $\Delta \Gc_{\mb{q}}=(\Vc_{\mb{q}},\Delta \mathcal{E}_{\mb{q}},\Delta \mathcal{A}_\mb{q}) $, that specifies connected graph component(s) corresponding to added/removed communication link(s) in the communication network such that
\begin{align}\label{eq:app_components_sets_1}
 \Vc_{\mb{q}} &= (\cup_{{\rm c}=1}^{\boldsymbol{\rm c}} \Dc_{\rm c}) \cup  \Dc_{\rm s},  \;\;\text{\rm s.t}\;\; \Vc_{\mb{q}} = \Vc,
 \\
    (i,j) &\in \Delta \mathcal{E}_{\mb{q}} \;\;\text{if}\;\; [\Delta \mathcal{A}_\mb{q}]_{i,j} =   a^{\mb{q}}_{ij}-a^{1}_{ij} \neq 0 \iff 
    \nonumber \\ \label{eq:app_components_sets_2}
   & \hspace{100pt}  [\Delta \Lc_{\mb{q}}]_{i,j} \neq 0,
\end{align}
where $\Dc_{\rm c}$ denotes the set of nodes (agents involved in switching links) in ${\rm c}$-th connected component with $|\Dc_{\rm c}| \geq 2$ and $\Dc_{\rm i'} \cap \Dc_{\rm j'} = \emptyset$ for any  $i',j' \in \{1, \cdots, \boldsymbol{\rm c}\},\, i'\neq j'$. Also, $\Dc_{\rm s}$ denotes the set of singletons i.e. single nodes that are not involved in any switching link.
% i.e. nodes that are connected to no others.
Then, there exists a permutation matrix ${\rm P}, \, {\rm P}{\rm P}^{\top} =I$ to relabel the nodes and represent the Laplacian matrix $\Delta \Lc_{\mb{q}}$ in block diagonal form, (cf. \cite[Ch. 6.12]{newman2018networks}), as follows
\begin{align}\label{eq:graph_component_lap}
    {\rm P}\Delta \Lc_{\mb{q}}{\rm P}^{\top}
    &=
    \breve{\Lc}_\mb{q}
    \nonumber \\
    &= 
    \diag{ \{\Delta \Lc_{\mb{q}}(\Dc_{\rm 1}), \cdots, \Delta \Lc_{\mb{q}}(\Dc_{\boldsymbol{\rm c}}), \Delta \Lc_{\mb{q}}(\Dc_{\rm s}) \}},
\end{align}
where $\Delta \Lc_{\mb{q}}(\Dc_{\rm c})$ denotes the Laplacian matrix of the ${\rm c}$-th connected component and $\Delta \Lc_{\mb{q}}(\Dc_{\rm s})=0$.
% \end{IEEEproof}

\begin{lemma}\label{lemma:app_component_measur}
Consider system in \eqref{eq:cl_sys} with topology switching from normal mode $\sigt=1$ to a safe mode $\sigt=\mb{q} \in \Qc$ and the measurements set $\Mc$ in \eqref{eq:measurments}, and let $\Delta \Lc_{\mb{q}} = \Lc_{\mb{q}} - \Lc_{1}$ denote the difference of the Laplacian matrices in safe and normal mode. Then
%
% \begin{enumerate}
%     \item $\Delta \Lc_{\mb{q}}$ can be represented in block-diagonal form as
%     %
%     \begin{align}
%     \hspace{-1ex}
%     \breve{\Lc}_\mb{q} 
%     =
%      \diag{ \{\Delta \Lc_{\mb{q}}(\Dc_{\rm 1}), \cdots, \Delta \Lc_{\mb{q}}(\Dc_{\boldsymbol{\rm c}}), \Delta \Lc_{\mb{q}}(\Dc_{\rm s}) \}}, 
%     \end{align}
%     %
%     where $\!\Delta \Lc_{\mb{q}}(\Dc_{\rm c})\!$ denotes the Laplacian matrix of the ${\rm c}$-th connected graph component among $\!\boldsymbol{\rm c}\!$ connected components (cf. Definition \ref{def:graph_component}) with $\Dc_{\rm c}\!$ denoting the set of nodes (agents involved in connected switching links) corresponding to ${\rm c}$-th connected component, and $\!\Delta \Lc_{\mb{q}}(\Dc_{\rm s}) \!=\! 0$ denotes the Laplacian matrix of the singletons with no connections that are grouped in set $\!\Dc_{\rm s}$.
%     %
%     \item Under condition
% %
% \begin{align}\label{eq:cond_suff_0}
%     \Img (\Delta \Lc_{\mb{q}}) \cap  \ker{([\mb{C}^{\top}_{\rm x}\; \mb{C}^{\top}_{\rm v}]^{\top})} =\emptyset,
% \end{align}
% %
% we have every connected graph component has at least one globally monitored node (agent), that is
% % in other words, connected intra-cluster switching links are linked to (a) monitored agent(s). 
% \begin{align}\label{eq:component_measurements}
%     \Dc_{\rm c} \cap \Mc \neq \emptyset, \ \ \forall \, {\rm c} \in \{1, \cdots, \boldsymbol{{\rm c}}\}.
% \end{align}
% where $\mb{C}_{\rm x}$ and $\mb{C}_{\rm v}$ are diagonal elements of $\mb{C}$ in \eqref{eq:measurments}. %
% \end{enumerate}
%
under condition
\begin{align}\label{eq:cond_suff_0}
    \Img (\Delta \Lc_{\mb{q}}) \cap  \ker{([\mb{C}^{\top}_{\rm x}\; \mb{C}^{\top}_{\rm v}]^{\top})} =\emptyset,
\end{align}
every connected graph component has at least one globally monitored node (agent), that is
% in other words, connected intra-cluster switching links are linked to (a) monitored agent(s).
%
\begin{align}\label{eq:app_component_measurements}
    \Dc_{\rm c} \cap \Mc \neq \emptyset, \ \ \forall \, {\rm c} \in \{1, \cdots, \boldsymbol{{\rm c}}\} .
\end{align}
where $\mb{C}_{\rm x}$ and $\mb{C}_{\rm v}$ are diagonal elements of $\mb{C}$ in \eqref{eq:measurments} and $\Dc_{\rm c}$ denotes the set of nodes in ${\rm c}$-th connected component of $\Delta \Lc_{\mb{q}}$ as given in \eqref{eq:app_components_sets_1}.
\end{lemma}
\begin{IEEEproof}
We first show \eqref{eq:cond_suff_0} is invariant under permutation of  $\Delta \Lc_{\mb{q}}$ which is introduced in \eqref{eq:graph_component_lap} and accordingly permutation of $[\mb{C}^{\top}_{\rm x}\; \mb{C}^{\top}_{\rm v}]^{\top}$. To this end, from the definition of nullspace we have
\begin{align}\label{eq:Lemma_kernel}
    \ker \paren{\begin{bmatrix}
 \mb{C}_{\rm x}\\
 \mb{C}_{\rm v}
 \end{bmatrix} \Delta\Lc_{\mb{q}}} 
%  \nonumber \\
 = \setdef{{x} \in \real^{N}}{ \begin{bmatrix}
 \mb{C}_{\rm x}\\
 \mb{C}_{\rm v}
 \end{bmatrix} \Delta\Lc_{\mb{q}}x = 0},
\end{align}
from which we obtain either
\begin{align}\label{eq:former}
 \Delta \Lc_{\mb{q}}{x} \notin \Img{(\Delta \Lc_{\mb{q}})} &\iff 
 \Delta \Lc_{\mb{q}}{x}=\boldsymbol{0}, 
\end{align}
or
\begin{align}\label{eq:latter}
\boldsymbol{0} \neq \boldsymbol{\Yc} = \Delta \Lc_{\mb{q}}{x} \in \Img{(\Delta \Lc_{\mb{q}})} &\implies 
 \begin{bmatrix}
 \mb{C}_{\rm x}\\
 \mb{C}_{\rm v}
 \end{bmatrix}  \boldsymbol{\Yc}=\boldsymbol{0}, 
\end{align} 
where the latter, \eqref{eq:latter}, is in contradiction with condition \eqref{eq:cond_suff_0}.
Now under the permutation defined in \eqref{eq:graph_component_lap}, $\begin{bmatrix}
 \mb{C}_{\rm x}\\
 \mb{C}_{\rm v}
 \end{bmatrix} \Delta\Lc_{\mb{q}}x = 0$ in \eqref{eq:Lemma_kernel} can be rewritten in block-partitioned diagonal form as
\begin{align}\label{eq:H_q_elements_stack_relabeled}
\begin{bmatrix}
 \mb{C}_{\rm x}\\
 \mb{C}_{\rm v}
 \end{bmatrix}
 {\rm P}^{\top}
\breve{\Lc}_\mb{q}
 {\rm P} x
 = 
 \begin{bmatrix}
 \mb{C}_{\rm x}\\
 \mb{C}_{\rm v}
 \end{bmatrix}
 {\rm P}^{\top}
\breve{\Lc}_\mb{q}
 \chi
 = 
 \begin{bmatrix}
 \breve{\mb{C}}_{\rm x}\\
 \breve{\mb{C}}_{\rm v}
 \end{bmatrix}
\breve{\Lc}_\mb{q}
 \chi
 =
 \boldsymbol{0}, 
\end{align}
in which $\chi={\rm P}x$ denotes the relabeled $x$ such that
\begin{align}\label{eq:app_states_relabeled}
    \chi&=\col{(\chi_1, \dots , \chi_{\boldsymbol{\rm c}})} ={\rm P}x, \;\, \text{with}
    \nonumber \\
    \chi_{\rm c} &= \col{(x_i)},\ \ \forall \, i \in \Dc_{\rm c},\ \ \forall \, {\rm c} \in \{1, \cdots, \boldsymbol{{\rm c}}\}.
\end{align}
Also, $ \breve{\mb{C}}_{\rm k} = \mb{C}_{\rm k} {\rm P}^{\top} = \begin{bmatrix}\mb{C}^{\rm 1}_{\rm k} & \cdots & \mb{C}^{\boldsymbol{\rm c}}_{\rm k}\end{bmatrix}$, ${\rm k} \in \{\rm x, v\}$ is a block-partitioned binary matrix that specifies monitored agents of each component \footnote{Note that ${\rm P}^{\top}$ permutes the columns of binary matrix $\mb{C}_{\rm k}$ whose row-vector elements are $\mathfrak{e}^{\top}_i,\, \forall \, i \in \Mc_{\rm k},\, {\rm k} \in \{\rm x, v\}$.}. 
%
% \begin{align}\label{eq:Lemma_kernel}
%     \ker \paren{\begin{bmatrix}
%  \mb{C}_{\rm x}\\
%  \mb{C}_{\rm v}
%  \end{bmatrix} \Delta\Lc_{\mb{q}}} 
% %  \nonumber \\
%  &= \setdef{{x} \in \real^{N}}{ 
%  \begin{bmatrix}
%  \mb{C}_{\rm x}\\
%  \mb{C}_{\rm v}
%  \end{bmatrix}
%  {\rm P}^{\top}
% \breve{\Lc}_\mb{q}
%  {\rm P} x = 0} 
%  %
%  \nonumber \\
%  &= \setdef{{x} \in \real^{N}}{ 
%  \begin{bmatrix}
%  \mb{C}_{\rm x}\\
%  \mb{C}_{\rm v}
%  \end{bmatrix}
%  {\rm P}^{\top}
% \breve{\Lc}_\mb{q}
%  \chi = 0} 
%  %
%  \nonumber \\
%  &= \setdef{{x} \in \real^{N}}{ 
%  \begin{bmatrix}
%  \breve{\mb{C}}_{\rm x}\\
%  \breve{\mb{C}}_{\rm v}
%  \end{bmatrix}
% \breve{\Lc}_\mb{q} \chi = 0, \chi = {\rm P}x} 
% %
% \nonumber \\
%  &= \setdef{\chi \in \real^{N}}{ 
%  \begin{bmatrix}
%  \breve{\mb{C}}_{\rm x}\\
%  \breve{\mb{C}}_{\rm v}
%  \end{bmatrix}
% \breve{\Lc}_\mb{q} \chi = 0, \chi = {\rm P}x} 
% %
% \nonumber \\
%  &= {\rm P}^{\top} 
%  \ker \paren{\begin{bmatrix}
%  \breve{\mb{C}}_{\rm x}\\
%  \breve{\mb{C}}_{\rm v}
%  \end{bmatrix} \breve{\Lc}_{\mb{q}}} 
% \end{align}
%
To show the results in \eqref{eq:former} and \eqref{eq:latter} hold also for the transformed form in \eqref{eq:H_q_elements_stack_relabeled}, one need to verify the invariance of \eqref{eq:cond_suff_0} under the permutation by ${\rm P}$, that is
\begin{align}\label{eq:cond_suff_0_invariant}
    \Img (\Delta \Lc_{\mb{q}}) \cap  \ker{([\mb{C}^{\top}_{\rm x}\; \mb{C}^{\top}_{\rm v}]^{\top})} &=\emptyset
    \iff
    \nonumber \\
    \Img (\breve{\Lc}_{\mb{q}}) \cap  \ker{([\breve{\mb{C}}^{\top}_{\rm x}\; \breve{\mb{C}}^{\top}_{\rm v}]^{\top})} &=\emptyset.
\end{align}
To show this, from the range and nullspace definition, for subspaces in \eqref{eq:cond_suff_0_invariant} we have
\begin{align}\label{eq:app_Img_Lap}
    \Img(\Delta \Lc_{\mb{q}}) 
    &= \setdef{\boldsymbol{\Yc} \in \real^{N}}{\boldsymbol{\Yc} = \Delta \Lc_{\mb{q}} x(t)},
    \\
    \ker \paren{\begin{bmatrix}
 \mb{C}_{\rm x}\\
 \mb{C}_{\rm v}
 \end{bmatrix}} 
 &=
 \ker(
 \mb{C}_{\rm x})
 \cap
 \ker(
 \mb{C}_{\rm v})
 \nonumber \\
 &= \setdef{\boldsymbol{\Xc} \in \real^{N}}{ \mb{C}_{\rm x} \boldsymbol{\Xc} = 0,\, \mb{C}_{\rm v} \boldsymbol{\Xc} = 0 },
\end{align}
and
\begin{align}
    \Img(\breve{\Lc}_{\mb{q}}) 
    &= 
    \setdef{ \breve{\boldsymbol{\Yc}} \in \real^{N}}{\breve{\boldsymbol{\Yc}}
    % ={\rm P}\boldsymbol{\Yc} 
    = \breve{\Lc}_{\mb{q}} \chi(t) = \breve{\Lc}_{\mb{q}}{\rm P}x(t)}
   \nonumber \\
  &= 
  \setdef{ \breve{\boldsymbol{\Yc}} \in \real^{N}}{{\rm P}^{\top}\breve{\boldsymbol{\Yc}}
    % ={\rm P}\boldsymbol{\Yc} 
    = {\rm P}^{\top}\breve{\Lc}_{\mb{q}} {\rm P}x(t) = \boldsymbol{\Yc} }
   \nonumber \\
 &= {\rm P}\, \Img{(\Delta \Lc_{\mb{q}})},
    \end{align}
   where we used \eqref{eq:graph_component_lap} and $\chi(t)={\rm P}x(t)$ as in \eqref{eq:app_states_relabeled} and \eqref{eq:app_Img_Lap}. Similarly,
    \begin{align}
    \ker \paren{\begin{bmatrix}
 \breve{\mb{C}}_{\rm x}\\
 \breve{\mb{C}}_{\rm v}
 \end{bmatrix}} 
 &=
 \ker(
 \breve{\mb{C}}_{\rm x})
 \cap
 \ker(
 \breve{\mb{C}}_{\rm v})
 \nonumber \\
 &= \setdef{\breve{\boldsymbol{\Xc}} \in \real^{N}}{ \breve{\mb{C}}_{\rm x} \breve{\boldsymbol{\Xc}} = 0,\, \breve{\mb{C}}_{\rm v} \breve{\boldsymbol{\Xc}} = 0 }
 \nonumber \\
 &= \setdef{\breve{\boldsymbol{\Xc}} \in \real^{N}}{ {\mb{C}}_{\rm x}{\rm P}^{\top} \breve{\boldsymbol{\Xc}} = 0,\, {\mb{C}}_{\rm v}{\rm P}^{\top} \breve{\boldsymbol{{\Xc}}} = 0 }
  \nonumber \\
 &= \setdef{\breve{\boldsymbol{\Xc}} \in \real^{N}}{ {\mb{C}}_{\rm x}{\boldsymbol{\Xc}} = 0,\, {\mb{C}}_{\rm v}{\boldsymbol{{\Xc}}} = 0,\, {\rm P} \boldsymbol{\Xc}=\breve{\boldsymbol{\Xc}}}
 \nonumber \\
 &= {\rm P}\, \ker \paren{\begin{bmatrix}
 \mb{C}_{\rm x}\\
 \mb{C}_{\rm v}
 \end{bmatrix}}.
\end{align}
%
% where $\Xc$ and $\breve{\boldsymbol{\Xc}}$ are a
%
Then
\begin{align}
    \Img(\breve{\Lc}_{\mb{q}}) 
    \cap
    \ker \paren{\begin{bmatrix}
 \breve{\mb{C}}_{\rm x}\\
 \breve{\mb{C}}_{\rm v}
 \end{bmatrix}} 
%   \nonumber \\
   &=
  {\rm P}\, \Img{(\Delta \Lc_{\mb{q}})}
  \cap
  {\rm P}\, \ker \paren{\begin{bmatrix}
 \mb{C}_{\rm x}\\
 \mb{C}_{\rm v}
 \end{bmatrix}} 
  \nonumber \\
  &=
  {\rm P} \paren{ \Img{(\Delta \Lc_{\mb{q}})}
  \cap
   \ker \paren{\begin{bmatrix}
 \mb{C}_{\rm x}\\
 \mb{C}_{\rm v}
 \end{bmatrix}}} 
 \nonumber \\
&=
 {\rm P} \paren{ \emptyset} = \emptyset.
\end{align}
where we used fact 2.9.29 in \cite{bernstein2009matrix} and condition \ref{cond:detection_condition_suff_0}.

Now one can proof \eqref{eq:app_component_measurements} by contradiction. Assume \eqref{eq:app_component_measurements} does not hold, that is $\exists \, {\rm c'} \in \{1, \cdots, \boldsymbol{{\rm c}}\},$ s.t. $\Dc_{\rm c'} \cap \Mc = \emptyset$, under which
% Then under \eqref{eq:graph_component_lap} and \eqref{eq:app_states_relabeled}, 
we have the ${\rm c'}$-th block in \eqref{eq:H_q_elements_stack_relabeled} such that
\begin{align}\label{eq:app_contra}
     \begin{bmatrix}
 \breve{\mb{C}}^{\rm c'}_{\rm x}\\
 \breve{\mb{C}}^{\rm c'}_{\rm v}
 \end{bmatrix}
\Delta \Lc_{\mb{q}}(\Dc_{\rm c'})
 \chi_{\rm c'}(t) = \boldsymbol{0}, \ \ \breve{\mb{C}}^{\rm c'}_{\rm x} = \breve{\mb{C}}^{\rm c'}_{\rm x} = 0, 
\end{align}
which holds for all $\chi_{\rm c'}(t)$ with $\Delta \Lc_{\mb{q}}(\Dc_{\rm c'})
 \chi_{\rm c'}(t) \in \Img{(\Delta \Lc_{\mb{q}}(\Dc_{\rm c'})) } \subseteq \Img{(\breve{\Lc}_{\mb{q}}) }$ as in \eqref{eq:app_contra}  $\Img{(\Delta \Lc_{\mb{q}}(\Dc_{\rm c'})) }  \in \ker{\paren{\begin{bmatrix}
 \breve{\mb{C}}^{\rm c'}_{\rm x}\\
 \breve{\mb{C}}^{\rm c'}_{\rm v}
 \end{bmatrix}}} 
  \implies 
 \Img{(\breve{\Lc}_{\mb{q}} )}  \cap \ker{([\breve{\mb{C}}^{\top}_{\rm x}\; \breve{\mb{C}}^{\top}_{\rm v}]^{\top})} \neq \emptyset$
that contradicts \eqref{eq:cond_suff_0_invariant}.  
\end{IEEEproof}
%%%%%%%%%%%%%%%%%%%%%%%%%%%%%%%%%%%%%%%%%%%%%%%%%%%%%%%%%%%%%%%%%%%%%%%%%%%%%%%%
\section{Proof of Lemma \ref{lemma:Inv_unobs_subspace}}\label{app_Inv_unobs_subspace}
Note that the Laplacian matrix $\Lc_{\sigt}$ of every connected undirected (or strongly connected and balanced directed) graph has only one zero eigenvalue, $\lambda = 0$, with the corresponding eigenvector $\boldsymbol{1}_N$ such that $\Lc_{\sigt}\boldsymbol{1}_{N} = \boldsymbol{0}$ \cite{olfati2004consensus}. Then, given the structure of $\mb{A}_{\sigt}$ in \eqref{eq:cl_sys},
% $\begin{bsmallmatrix} {\scriptscriptstyle (1/\sqrt{N})}\boldsymbol{1}^{\top}_{\s \! N} & \boldsymbol{0}^{\top}_{\!N} \end{bsmallmatrix}^{\scriptscriptstyle{\top}}$. Then
$(\lambda = 0, w_r =\begin{bsmallmatrix} {\scriptscriptstyle (1/\sqrt{N})}\boldsymbol{1}_{\s \! N} \\ \boldsymbol{0}_{\!N} \end{bsmallmatrix})$ is an eigenpair of system matrix $\mb{A}_{\sigt}$ associated with that of Laplacian $\Lc_{\sigt}$ with $\sigma(t_{k-1})=\mb{q} \in \Qc, \, \, t\in [t_{k-1}, t_k)$. Also, 
it can be verified that the eigenpair $(\lambda = 0, w_r)$ lies in the unobservable subspace of system \eqref{eq:cl_sys} as it is a nontrivial solution to the PBH test for observability:
\begin{align}\label{eq:PBH_test}
    \begin{bmatrix}
        \lambda I_{} -\mb{A}_{\mb{q}} \\ \mb{C}_{} 
    \end{bmatrix}
    w_r
    &=
    \boldsymbol{0}  , \quad  \, \lambda = 0 \in \cplx,\\
    \mb{C}_{} &= \diag\braces{0, C_{\rm v}}.
\end{align}
%
% \begin{align}\label{eq:}
% \lambda \mb{x}_p - \mb{x}_v &= \boldsymbol{0} \\
% %
% \Lc_{k} \mb{x}_p + (\lambda + 1) \mb{x}_v &= \boldsymbol{0},\\
% \mb{C}w_r = \diag\braces{0, C_v} w_r &= \boldsymbol{0}
% \end{align}.
%
Therefore, one can conclude that the right eigenvector $w_r$ contained in $\ker(\mb{C}) $ belongs to $\ker(\Oc_k)$ that is defined in \eqref{eq:obsv_nullspace_matrix} \cite[Th. 15.8]{hespanha2018linear}. 
Furthermore, as $(\lambda = 0, w_r)$ is the eigenpair associated with the equilibrium subspace \eqref{eq:cond_consensus} of every $\mb{A}_{\mb{q}}$ with Laplacian $\Lc_{\mb{q}}$, it is straightforward from Lemma \ref{lemma:observability} that $ \spann \braces{w_r} = \spann \braces{ \begin{bsmallmatrix} \boldsymbol{1}_{\! N} \\ \boldsymbol{0}_{\!N} \end{bsmallmatrix}} \subseteq  \boldsymbol{\Nc}_1^{\infty} = \ker(\boldsymbol{\Oc}) $ over $t \in [t_0, +\infty)$.
%
%%%%%%%%%%%%%%%%%%%%%%%%%%%%%%%%%%%%%%%%%%%%%%%%%%%%%%%%%%%%%%%%%%%%%%%%%%%%%%%%
\section{Proof of Proposition~\ref{prop:attack_detectability_local}}\label{app_attack_detectability_local}
Let $\sigt=\mb{q} \in \Qc, \, \, t\in [t_{k-1}, t_k) $ and consider the error dynamics of local observers in \eqref{eq:obs_decent_error}. According to Definition \ref{def:inv_zeros}, a ZDA for \eqref{eq:obs_decent_error} should satisfy
\begin{equation}\label{eq:obs_decent_INV_zero_1}
    \begin{bmatrix}
        % \lambda_0 I_{2N_i} - (\bar{\mb{A}}^i_{\sigt}-\bar{\mb{K}}_{\sigt}\mb{C}_i) 
        \lambda_0 I_{} - {\mb{F}}^{{\rm i}}_{\mb{q}}
        & -\mb{T}_{}^{\rm i} \mb{B}^{\rm i} \\ \mb{C}_{{\rm i}_i} & 0
    \end{bmatrix}
    \begin{bmatrix}
        \tilde{\mb{e}}_{\rm i}(0) \\ \mb{u}_{0_{\rm i}}
    \end{bmatrix}
    =
    \begin{bmatrix}
        \boldsymbol{0} \\ \boldsymbol{0}
    \end{bmatrix},
\end{equation}
where $\tilde{\mb{e}}_{\rm i}(0) := \mb{e}_{\rm i}(0)-\bar{\mb{e}}_{\rm i}(0)=\tilde{\mb{x}}_{0_{\rm i}}$. 
Also, by considering \eqref{eq:uio_conds_2} and the fact that $ \mb{C}_{{\rm i}_i} \tilde{\mb{e}}_{\rm i}(0)=\mb{C}_{{\rm i}_i} \tilde{\mb{x}}_{\rm i}(0)=\boldsymbol{0}$ in the second equation of \eqref{eq:obs_decent_INV_zero_1}, matrix pencil \eqref{eq:obs_decent_INV_zero_1}  can be rewritten as
\begin{equation}\label{eq:obs_decent_INV_zero_2}
\small
\underbrace{
    \begin{bmatrix}
        \lambda_0 I_{} - {\bar{\mb{A}}^{\rm i}_{\mb{q}} }
        & - \mb{T}_{}^{\rm i}\mb{B}^{\rm i} \\ \mb{C}_{{\rm i}_i} & 0
    \end{bmatrix} }_{\bar{\mb{P}}}
    \begin{bmatrix}
        \tilde{\mb{x}}_{\rm i}(0) \\ \mb{u}_{0_{\rm i}}
    \end{bmatrix}
    =
    \begin{bmatrix}
        \boldsymbol{0} \\ \boldsymbol{0}
    \end{bmatrix}.
\end{equation}
It is immediate from Definition \ref{def:inv_zeros} that stealthy attack $\mb{u}_{a_{\rm i}}$ in \eqref{eq:obs_decent_error}, in the both ZDA and covert attack, looses its stealthiness with respect to the local residual $\mb{r}_{{\rm i}_i}$ if, and only if, there is no non-trivial zeroing direction associated with matrix pencil in \eqref{eq:obs_decent_INV_zero_1} or equivalently $\bar{\mb{P}}$ in \eqref{eq:obs_decent_INV_zero_2}, which in turn implies $\bar{\mb{P}}$ has full rank.
Moreover, from Definition \ref{def:inv_zeros} and condition \eqref{eq:E_definition}, it is straightforward that matrix pencil $\mb{P}$,  defined in \eqref{eq:pencil_uio}, is associated with the zeroing direction of the local system \eqref{eq:cluster_dyn2}.
We now show how conditions ({i})-({iii})  establish the equivalence of rank sufficiency for $\mb{P}$ in \eqref{eq:pencil_uio} and $\bar{\mb{P}}$ in \eqref{eq:obs_decent_INV_zero_2}.
Given $\mb{P}$ in \eqref{eq:pencil_uio}, one can write
\begin{align}\label{eq:pencil_uio_pre} 
    & \begin{bmatrix}
         I_{} - {\mb{h}}^{\rm i}_{} \mb{C}_{{\rm i}_i} & \lambda_0 {\mb{h}}^{\rm i}_{}
        \\
        0 & I_{} \\
        {\mb{h}}^{\rm i}_{}  \mb{C}_{{\rm i}_i} & - \lambda_0 {\mb{h}}^{\rm i}_{}
    \end{bmatrix} 
    \mb{P} = \nonumber \\
   & 
    \begin{bmatrix}
        \lambda_0 I_{} -\bar{\mb{A}}^{\rm i}_{\mb{q}} & -(I_{} - {\mb{h}}^{\rm i}_{} \mb{C}_{{\rm i}_i}){\mb{B}}^{\rm i} & 0
        \\
        \mb{C}_{{\rm i}_i} & 0 & 0 \\
        - {\mb{h}}^{\rm i}_{}  \mb{C}_{{\rm i}_i} {\mb{A}}^{\rm i}_{\mb{q}}  & {\mb{h}}^{\rm i}_{} \mb{C}_{{\rm i}_i}{\mb{B}}^{\rm i} & {\mb{E}}^{\rm i}_{}
    \end{bmatrix},
\end{align}
where ${\mb{h}}^{\rm i}_{} :=  {\mb{E}}^{\rm i}_{} (\mb{C}_{{\rm i}_i}{\mb{E}}^{\rm i}_{})^{\dagger}$ is a solution to \eqref{eq:uio_conds_1} that exists under condition ({ii}) \cite[Lemma 1]{chen1996design}. Then, postmultiplying \eqref{eq:pencil_uio_pre} by
\begin{align}\label{eq:pencil_uio_post}
    \begin{bmatrix}
     I_{} & 0 & 0
        \\
        0 & I_{} & 0 \\
        (\mb{C}_{{\rm i}_i}{\mb{E}}^{\rm i}_{})^{\dagger}\mb{C}_{{\rm i}_i}\mb{A}^{\rm i}_{\mb{q}}  & (\mb{C}_{{\rm i}_i}{\mb{E}}^{\rm i}_{})^{\dagger}\mb{C}_{{\rm i}_i}\mb{B}^{\rm i} & I_{}
    \end{bmatrix},
\end{align}
and considering \eqref{eq:uio_conds_2} yields
\begin{align}
    \begin{bmatrix}
        \lambda_0 I_{} -\bar{\mb{A}}^{\rm i}_{\mb{q}} & -{\mb{T}}^{\rm i}_{} {\mb{B}}^{\rm i} & 0
        \\
        \mb{C}_{{\rm i}_i} & 0 & 0 \\
        0  & 0 & {\mb{E}}^{\rm i}_{}
    \end{bmatrix}.
\end{align}
Since node $i \in \Pc_{\rm i}$ is $\mb{k}$-connected, we have $|\Nc_{i}|=\mb{k}$ and $\mb{k}\leq \rank \paren{ \mb{C}_{{\rm i}_i} } \leq 2\mb{k}$ (cf. \eqref{eq:measurments}). Then, from condition ({i}), one can verify that $\rank \paren{ \mb{C}_{{\rm i}_i} } \geq  \rank \paren{ \mb{B}^{\rm i} } + \rank \paren{ \mb{E}^{\rm i}_{} }$ guarantees \eqref{eq:pencil_uio} is a tall or square matrix pencil having only a finite number\footnote{This condition is not valid for degenerate systems which are out of scope of this work.} of output-zeroing directions \cite[Ch. 2]{lee2017l1}. Also, the pre- and post-multiplied matrices in \eqref{eq:pencil_uio_pre} and \eqref{eq:pencil_uio_post} are full column rank. Therefore, we have %
\begin{align}\label{eq:pencil_equivalance}
\rank \paren{ \mb{P}}
= 
\rank
\underbrace{
    \begin{bmatrix}
        \lambda_0 I_{} -\bar{\mb{A}}^{\rm i}_{\mb{q}} & -{\mb{T}}^{\rm i}_{} {\mb{B}}^{\rm i}_{} 
        \\
        \mb{C}_{{\rm i}_i} & 0 
    \end{bmatrix}}_{\bar{\mb{P}}}
    + 
\rank \paren{ {\mb{E}}^{\rm i}_{}}.
\end{align}
Recall ${\mb{E}}^{\rm i}_{}$ is full column rank, and hence ${\mb{P}}$ in \eqref{eq:pencil_uio} is full rank if, and only if, $\bar{\mb{P}}$ in \eqref{eq:pencil_equivalance} is full rank. This guarantees that a locally undetectable stealthy attack is impossible.
%
%%%%%%%%%%%%%%%%%%%%%%%%%%%%%%%%%%%%%%%%%%%%%%%%%%%%%%%%%%%%%%%%%%%%%%%%
%%%%%%%%%%%%%%%%%%%%%%%%%%%%%%%%%%%%%%%%%%%%%%%%%%%%%%%%%%%%%%%%%%%%%%%%
\section{Proof of Theorem~\ref{th:Attack_detectability_switching}}\label{app_Attack_detectability_switching}
Consider \eqref{eq:obs_cent_error} over $ t\in [t_0,+\infty)$, and let the safe mode $ \sigt \!=\! \mb{q} \! \in \!  \Qc,$ $ t \in [t_1, +\infty)$ the continuous system residual $\mb{r}_0(t)$ and its successive derivatives can be rewritten as
%
% {\small
\begin{align}\label{eq:Y}
    \mb{R} &= 
    \mb{\Oc}_1 \mb{e}(t) - \boldsymbol{\Hc}_{}( \mb{H}\mb{C}) \mb{E}
    +
    \boldsymbol{\Hc}(\mb{B}) \mb{U}_a
    +
    \boldsymbol{\Hc}_{}( \mb{H}) \mb{U}_s
    \nonumber \\
    & 
    \qquad \qquad \qquad \qquad \quad \ \ \:{-}\:
    \mb{U}_s 
    +
    \boldsymbol{\Hc}_{}( \Delta \mb{A}_{\mb{q}}) \mb{X},
\end{align}
% }
%
where 
% \begin{align}\label{eq:output_sequence}
% %
% \mb{R} &= 
% \col\paren{
%     \mb{r}_0(t),\:\, \dot{\mb{r}}_0(t),\:\, \cdots,\:\, {\mb{r}_0^{ (2N-1)}(t)}
% },\\
% %
% \mb{U}_{\rm j} &= 
% \col\paren{
%     \mb{u}_{{\rm{j}}}(t),\:\,  \dot{\mb{u}}_{{\rm j}}(t),\,\, \cdots,\:\, {\mb{u}_{{\rm j}}^{(2N-1)}(t)}
% },\\
% \mb{E} &= 
% \col\paren{
%     \mb{e}(t),\:\: \dot{\mb{e}}(t),\:\: \cdots,\:\: {\mb{e}^{(2N-1)}(t)}
% },\\ 
% %
% \mb{X} &= 
% \col\paren{
%     \mb{x}(t),\:\: \dot{\mb{x}}(t),\:\: \cdots,\:\: {\mb{x}^{(2N-1)}(t)}
% },\\ \label{eq:H_q}
% %
% \boldsymbol{\Hc}_{}({b})&= 
% \begin{bmatrix}
%     0   & 0 & 0 & \cdots & 0 \\
%     \mb{C} {b} & 0 & 0 & \cdots& 0 \\
%     \mb{C}{\mb{A}}_{1} {b} & \mb{C} {b} & 0 & \cdots& 0 \\
%     \vdots & \vdots & \ddots & \ddots & \vdots\\
%     \mb{C}{\mb{A}}_{1}^{2N-2} {b} & \mb{C}{\mb{A}}_{1}^{2N-3} {b} & \cdots & \mb{C} {b} & 0 
% \end{bmatrix},
% \end{align}
%%
% different notation than \col
\begin{align}\label{eq:output_sequence}
\mb{R} &= 
\begin{bmatrix}
    \mb{r}^{\top}_0(t)& \dot{\mb{r}}^{\top}_0(t) & \cdots & {(\mb{r}_0^{\top}(t))^{({\rm d})}}
\end{bmatrix}^{\top},\\
\mb{U}_{\jmath} &= 
\begin{bmatrix}
    \mb{u}^{\top}_{{\jmath}}(t) & \dot{\mb{u}}^{\top}_{{\jmath}}(t) & \cdots & ({\mb{u}^{\top}_{{\jmath}}(t))^{({\rm d})}}
\end{bmatrix}^{\top},\\ \label{eq:output_sequence_E}
\mb{E} &= 
\begin{bmatrix}
    \mb{e}^{\top}(t) & \dot{\mb{e}}^{\top}(t) & \cdots & {(\mb{e}^{\top}(t))^{({\rm d})}}
\end{bmatrix}^{\top},\\ \label{eq:output_sequence_X}
\mb{X} &= 
\begin{bmatrix}
    \mb{x}^{\top}(t) & \dot{\mb{x}}^{\top}(t) & \cdots & {(\mb{x}^{\top}(t))^{({\rm d})}}
\end{bmatrix}^{\top},\\ \label{eq:H_q}
\boldsymbol{\Hc}_{}({b})&= 
\begin{bmatrix}
    0   & 0 & 0 & \cdots & 0 \\
    \mb{C} {b} & 0 & 0 & \cdots& 0 \\
    \mb{C}{\mb{A}}_{1} {b} & \mb{C} {b} & 0 & \cdots& 0 \\
    \vdots & \vdots & \ddots & \ddots & \vdots\\
    \mb{C}{\mb{A}}_{1}^{{\rm d}} {b} & \mb{C}{\mb{A}}_{1}^{{\rm d}-1} {b} & \cdots & \mb{C} {b} & 0 
\end{bmatrix},
\end{align}
with $ {\jmath}\in \braces{a,s}$, $ {b}\in\braces{\mb{B},\mb{H}\mb{C},\mb{H},\Delta \mb{A}_{\mb{q}}}$, $\Delta \mb{A}_{\mb{q}} = (\mb{A}_{\mb{q}}-\mb{A}_{1}) $ and ${\rm d} \in \naturals \setminus \{1,2\}$.

From \eqref{eq:stealthy} in Proposition \ref{propo:stealthy_attacks} and \eqref{eq:est_error_normal}-\eqref{eq:est_error}, it can be easily verified that \eqref{eq:Y} is simplified to $\mb{R}=\mb{\Oc}_1 \bar{\mb{e}}(t) - \boldsymbol{\Hc}_{}( \mb{H}\mb{C}) \bar{\mb{E}} + \boldsymbol{\Hc}_{}( \Delta \mb{A}_{\mb{q}}) \mb{X}$ where $ \bar{\mb{E}}$ has the same form as \eqref{eq:output_sequence_E} while whose elements are $\bar{\mb{e}}$ and its derivatives. Therefore, in a stealthy attack case  $\lim_{t_1 \rightarrow \infty} \mb{R} = \boldsymbol{0}$ during normal mode over $ t\in [t_0,t_1)$. The objective is to characterize the effect of switching communication, modeled as discrepancy $\Delta \mb{A}_{\mb{q}}$ in \eqref{eq:obs_cent_error} and \eqref{eq:Y}, on the stealthiness of attacks in the residual $\mb{r}_0(t)$ of centralized observer \eqref{eq:obs_cent} during safe mode over $ t\in [t_1, +\infty) $ (cf. Problem \ref{prob:attack_detection}). 
% To this end, the proof follows the notion of input observability \cite{massoumnia1989failure} and the fundamental problem of residual generation \cite{de2001geometric} for input directions spanned by $\Img( \Delta \mb{A}_{\mb{q}})$ and their effects in residual $\mb{r}_0(t)$.
%
% We analyze the first topology switching to a safe mode over $ t \in [t_1, t_2 \rightarrow +\infty)$ and the same analysis is applied to further topology switching. 
Given the input-output matrix \eqref{eq:H_q} for the switching perturbations $\Delta \mb{A}_{\mb{q}}$ in \eqref{eq:Y}, note that $\boldsymbol{\Hc}_{}( \Delta \mb{A}_{\mb{q}}) \mb{X} = \boldsymbol{0} $ over $ t\in [t_1,+\infty)$ in \eqref{eq:Y} is
the necessary condition under which the stealthy attacks, modeled in \eqref{eq:attack_model}, remain undetectable in the residual $\mb{r}_0(t)$ of \eqref{eq:obs_cent_error}, regardless of the perturbation $\Delta \mb{A}_{\mb{q}} \mb{x}$ caused by topology switching. Therefore, $\boldsymbol{\Hc}_{}( \Delta \mb{A}_{\mb{q}}) \mb{X} \neq \boldsymbol{0} $ in \eqref{eq:Y} implying the system switching $\Delta \mb{A}_{\mb{q}}$ affects $\mb{R}(t),\,  t\in[t_1,+\infty)$ in \eqref{eq:Y} guarantees attack detectability in $\mb{r}_0(t)$.

Consider Markov parameters $\mb{C}{\mb{A}}_{1}^{\rm{d}}\Delta \mb{A}_{\mb{q}},$ $ \rm{d} \in \naturals_{0}$ in \eqref{eq:H_q}, the term $\boldsymbol{\Hc}_{}( \Delta \mb{A}_{\mb{q}}) \mb{X} $ in \eqref{eq:Y} can be rewritten as
\begin{align}\label{eq:H_q_elements}
\sum\limits_{ l = 0}^{\rm d}
    \mb{C}{\mb{A}}_{1}^{{\rm{d}}}\Delta \mb{A}_{\mb{q}}\mb{x}^{({\rm d}-l)}(t) = \boldsymbol{0}, \ \ \forall \, {\rm d} \in \naturals_{0}.
\end{align}
We show that under condition \ref{cond:detection_condition_suff_0}, the first two terms in \eqref{eq:H_q_elements} are non-zero (and so is $\boldsymbol{\Hc}_{}( \Delta \mb{A}_{\mb{q}}) \mb{X} \neq \boldsymbol{0} $) unless $\Delta \mb{A}_{\mb{q}}\mb{x}(t)=\boldsymbol{0},\ \ \forall \,t \in [t_1,  +\infty)$.

By setting $\rm d = 0, 1$, and expanding \eqref{eq:H_q_elements} we obtain
\begin{align} \label{eq:H_q_elements_d0}
    {\rm d}=0 
    \overset{\text{\eqref{eq:H_q_elements}}}{\Rightarrow} \ \
    \mb{C}_{\rm v}\Delta\Lc_{\mb{q}}x(t)&=\boldsymbol{0}, 
    \ \ \forall \,t \in [t_1,  +\infty),
    \\ \label{eq:H_q_elements_d1}
    {\rm d}=1 
    \overset{\text{\eqref{eq:H_q_elements}}}{\Rightarrow} \ \
    \mb{C}_{\rm v}\Delta\Lc_{\mb{q}}v(t)&=\boldsymbol{0}, \ \ \text{and,} \nonumber \\ 
    \mb{C}_{\rm x}\Delta\Lc_{\mb{q}}x(t)&=\boldsymbol{0},
    \ \ \forall \,t \in [t_1,  +\infty),
\end{align}
% \begin{align}
%     {\rm d}=0,1 \Rightarrow \mb{C}_{\rm v}\Delta\Lc_{\mb{q}}x(t)=\boldsymbol{0}, \ \
%     \mb{C}_{\rm v}\Delta\Lc_{\mb{q}}v(t)=\boldsymbol{0},
% \end{align}
% and
% \begin{align}
%     {\rm d}=1,2 \Rightarrow \mb{C}_{\rm x}\Delta\Lc_{\mb{q}}x(t)=\boldsymbol{0}, \ \
%     \mb{C}_{\rm x}\Delta\Lc_{\mb{q}}v(t)=\boldsymbol{0},
% \end{align}
%
where $\mb{C}_{\rm x}$ and $\mb{C}_{\rm v}$ are diagonal elements of $\mb{C}$ as given in \eqref{eq:cl_sys}-\eqref{eq:measurments}, $\Delta \Lc_{\mb{q}} = \Lc_{\mb{q}} - \Lc_{1}$ is the non-zero submatrix of $\Delta \mb{A}_{\mb{q}} = (\mb{A}_{\sigt}-\mb{A}_{1})= \begin{bsmallmatrix}
    0_{} & 0 \\
    - \alpha \Delta \Lc_{\mb{q}} &  0
    \end{bsmallmatrix}$, and ${\mb{x}}(t) = \col({x}(t),{v}(t))$ as in \eqref{eq:cl_sys}.
Then, using \eqref{eq:H_q_elements_d0} and \eqref{eq:H_q_elements_d1}, we have
\begin{align}\label{eq:H_q_elements_stack}
     \begin{bmatrix}
 \mb{C}_{\rm x}\\
 \mb{C}_{\rm v}
 \end{bmatrix}
 \Delta\Lc_{\mb{q}}x(t)=\boldsymbol{0}, \ \ \forall \,t \in [t_1,  +\infty).
\end{align}
%
% Note that from condition \ref{cond:detection_condition_suff_0}, \eqref{eq:measurments} and the relation between the range of multiplied matrices\footnote{For given matrices ${\rm A} \in \real^{p \times n}$ and ${\rm B} \in  \real^{n \times m}$, $\Img ({\rm A B}) = {\rm A} \, \Img ({\rm B})$, \cite[P. 43]{poznyak2009advanced}.}, for $ \mb{C}_{\rm v}\Delta\Lc_{\mb{q}}$ and $ \mb{C}_{\rm x}\Delta\Lc_{\mb{q}}$ in \eqref{eq:H_q_elements_d0} and \eqref{eq:H_q_elements_d1} we have
%
%
% \begin{align}\label{eq:H_q_elements_Img}
%  \mathfrak{e}^{\top}_{i}\Delta\Lc_{\mb{q}} &= [\Delta\Lc_{\mb{q}}]_{i,:} \neq 0, \ \ \forall \, i  \in \Mc \implies 
%   \nonumber \\ 
%   \Img{(\mb{C}_{\rm v}\Delta\Lc_{\mb{q}})} &= \mb{C}_{\rm v}\Img{(\Delta\Lc_{\mb{q}})} \subseteq \Img{(\Delta\Lc_{\mb{q}})}\neq 0, \ \ \text{and/or}
%   %
%   \nonumber \\
%   \Img{(\mb{C}_{\rm x}\Delta\Lc_{\mb{q}})} &= \mb{C}_{\rm x}\Img{(\Delta\Lc_{\mb{q}})} \subseteq \Img{(\Delta\Lc_{\mb{q}})} \neq 0,
% \end{align}
%  
%
%   
%
% with $\mathfrak{e}_{i}$ being the $i$-th canonical vector in  $\real^{N}$ that here, in the form of row elements in $\mb{C}_{\rm x}$ or $\mb{C}_{\rm v}$, specifies $i$-th globally monitored agent which is also involved in switching.
%
Under condition \ref{cond:detection_condition_suff_0}, one can verify that \eqref{eq:H_q_elements_stack} implies
\begin{align}\label{eq:H_q_elements_kernel}
 \Delta \Lc_{\mb{q}}{x}(t) \notin \Img{(\Delta \Lc_{\mb{q}})} \iff 
 \Delta \Lc_{\mb{q}}{x}(t)=\boldsymbol{0}, \ \ \forall \,  t \in [t_1, +\infty).
\end{align}
%
% \begin{align}\label{eq:H_q_elements_kernel}
%   \Delta \Lc_{\mb{q}}{x}(t)=\boldsymbol{0} \iff {x}(t) \in \ker{(\Delta \Lc_{\mb{q}})}, \ \ \forall \,  t \in [t_1, +\infty).
% \end{align}
%
otherwise, for any $x(t)$ such that $   \boldsymbol{0} \neq \Delta \Lc_{\mb{q}}{x}(t)= \boldsymbol{\Yc} \in \Img{(\Delta \Lc_{\mb{q}})} $, we obtain $[\mb{C}^{\top}_{\rm x}\; \mb{C}^{\top}_{\rm v}]^{\top} \boldsymbol{\Yc} = 0,\, \boldsymbol{\Yc} \in \ker{([\mb{C}^{\top}_{\rm x}\; \mb{C}^{\top}_{\rm v}]^{\top})}  $ for \eqref{eq:H_q_elements_stack}, which is in contradiction with condition \ref{cond:detection_condition_suff_0}.

Now considering the consensus protocol \eqref{eq:ctrl_proto}, it can be verified that $\Delta \Lc_{\mb{q}}$ (or equivalently $\Delta \mb{A}_{\mb{q}}$ in \eqref{eq:H_q_elements}), encodes connected graph component(s) corresponding to added/removed communication link(s) in the communication network (cf. Definition \ref{def:graph_component} and \ref{prop:app_def_diagonal}). Then an elementary transformation, by means of the permutation matrix ${\rm P}$ as defined in  \eqref{eq:graph_component_lap}, transforms \eqref{eq:H_q_elements_kernel} into block-diagonal form as
%
% \begin{align}\label{eq:H_q_elements_kernel_relabeled}
%     \Delta \Lc_{\mb{q}}{x}(t)
%     =
%     {\rm P}^{\top}
%     \breve{\Lc}_{\mb{q}}\chi(t)
%     =
%     \boldsymbol{0}, \ \ \forall \,  t \in [t_1, +\infty),
% \end{align}
%
\begin{align}\label{eq:H_q_elements_kernel_relabeled}
    \Delta \Lc_{\mb{q}}{x}(t)
    =
    \boldsymbol{0}
    \iff
    \breve{\Lc}_{\mb{q}}\chi(t)
    =
    \boldsymbol{0},
    \ \ \forall \,  t \in [t_1, +\infty),
\end{align}
where the block-diagonal $\breve{\Lc}_{\mb{q}}$ is given in \eqref{eq:graph_component_lap} and $\chi(t)={\rm P}x(t)$ denotes the relabeled system states such that
\begin{align}\label{eq:states_relabeled}
    \chi(t)&=\col{(\chi_1(t), \dots , \chi_{\boldsymbol{\rm c}}(t))} ={\rm P}x(t), \;\, \text{with}
    \nonumber \\
    \chi_{\rm c}(t) &= \col{(x_i(t))},\ \ \forall \, i \in \Dc_{\rm c},\ \ \forall \, {\rm c} \in \{1, \cdots, \boldsymbol{{\rm c}}\},
\end{align}
with $\Dc_{\rm c}$ being the set of nodes (agents involved in switching links) in ${\rm c}$-th connected component\footnote{Although the analysis here is at the global level, it is worth mentioning that $\Delta \Lc_{\mb{q}}$ at cluster levels i.e. $\Pc_{\rm i},\, {\rm i} \in \braces{1,\cdots,|\boldsymbol{\Pc}|}$ may have more than one connected component.} as in \eqref{eq:app_components_sets_1}. 
Also, note that permutation matrix ${\rm P}$ is a binary nonsingular matrix by definition, and that the Laplacian matrix is zero row sum matrix and, if connected, its nullspace is spanned by $\boldsymbol{1}$, a vector of all ones \cite{olfati2004consensus}. Therefore, from \eqref{eq:H_q_elements_kernel_relabeled} and for nodes participated in (connected) switching links, i.e. $ \forall \, i,j \in \Dc_{\rm c},\, i\neq j$, one can conclude that
\begin{align}\label{eq:H_q_elements_kernel_1}
    {x}_i^{}(t) - {x}_j^{}(t) &= 0 \Leftrightarrow
    \nonumber \\ 
    {x}_i^{}(t) &= {x}_j^{}(t), &\forall& \, i, j \in \Dc_{\rm c}, \ \ \forall\, {\rm c} \in \{1, \cdots, \boldsymbol{\rm c}\}, \nonumber \\
    &  &\forall& \, t \in [t_1, +\infty),
\end{align}
which by considering the continuity of the system states can be extended for its higher-order time derivatives and be rewritten as 
%
% \begin{align}\label{eq:H_q_elements_kernel_2_2}
%     {x}_i^{(\rm m)}(t) = {x}_j^{(\rm m)}(t), \ \ &\forall \, i, j \in \Dc_{\rm c}, \ \ {\rm c} \in \{1, \cdots, \boldsymbol{\rm c}\}, 
%     \nonumber \\
%     &\forall \, {\rm m} \in \naturals_0, \ \ \forall \,t \in [t_1, +\infty),
% \end{align}
%
\begin{align}\label{eq:H_q_elements_kernel_2_2}
    (\mathfrak{e}^{\top}_{i}-\mathfrak{e}^{\top}_{j}) {x}^{(\rm m)}(t) = 0, \ \ &\forall \, i, j \in \Dc_{\rm c}, \ \ 
    \forall \,{\rm c} \in \{1, \cdots, \boldsymbol{\rm c}\}, \nonumber \\
    &\forall \, {\rm m} \in \naturals_0, \ \ \forall \,t \in [t_1, +\infty),
\end{align}
with $\mathfrak{e}_{i},\,  \mathfrak{e}_{j}$ being $i$-th and $j$-th standard-basis vectors in $\real^{N}$.
%
% Consider \eqref{eq:H_q_elements_4} and let $\mathfrak{C}$ be a block-partitioned vector such that $ \mathfrak{C}=\col( \begin{bmatrix}
%     (\mathfrak{e}^{\top}_{i}-\mathfrak{e}^{\top}_{j}) & 0
% \end{bmatrix})$ for $\forall \,i, j \in \Dc_{\rm c}, \, {\rm c} \in \braces{1,\cdots,\boldsymbol{\rm c}}$. we equivalently obtain 
% %
% \begin{equation}\label{eq:H_q_elements_5}
%     \check{\mb{y}}^{({\rm m})} = \mathfrak{C} \mb{x}^{({\rm m})}=0, \, \forall \, {\rm m} \in \naturals_{0}, \, \forall \, t \in [t_1, \infty).
% \end{equation}
%

Also, from \eqref{eq:H_q_elements}, \eqref{eq:H_q_elements_kernel_relabeled}, \eqref{eq:H_q_elements_kernel_2_2} and by considering the structure $\mb{A}_{\sigt}$ and system state $\mb{x}(t) = \col(x(t),v(t))$ in \eqref{eq:cl_sys}, we obtain
\begin{align}\label{eq:H_q_elements_2}
    \Delta \mb{A}_{\mb{q}}\mb{x}^{(\rm m)}(t) &= \boldsymbol{0} 
    \Leftrightarrow \nonumber \\
    \Delta \Lc_{\mb{q}}{x}^{(\rm m)}(t) &= \boldsymbol{0}, \ \ \forall \,  {\rm m}\in \naturals_0,\, \forall \,  t \in [t_1, +\infty).
\end{align}
Therefore, under condition \ref{cond:detection_condition_suff_0}, one can conclude that unless \eqref{eq:H_q_elements_kernel}/\eqref{eq:H_q_elements_kernel_2_2} holds that is the system states (positions $x_i(t)$, $x_j(t)$ and their successive derivatives) of all agents within each graph component, i.e. agents involved in connected intra-cluster switching links, are respectively identical $\forall \, t \in [t_1, 
+\infty)$, the left side of \eqref{eq:H_q_elements_d0} and \eqref{eq:H_q_elements_d1} is non-zero and so is \eqref{eq:H_q_elements}, implying $\Delta \mb{A}_{\mb{q}}$ affects $\mb{R}(t)$ whereby the attacks are detectable. 

% We now consider the effect of attack input $\mb{B}\mb{u}_a$ on \eqref{eq:H_q_elements_4} in both zero-dynamics and covert attack, and show the limitations of attack detection by the central observer's residual $\mb{r}_0(t)$ using switching topology.

We now show under conditions \ref{cond:detection_condition_suff_1}-\ref{cond:detection_condition_suff_2} the domain of existence of \eqref{eq:H_q_elements_kernel} is shrank into the only case that the entire system states, except for those affected by stealthy attacks, are at an equilibrium. 
    
% Given this new output with $\dot{\mb{x}}$-dynamics in \eqref{eq:cl_sys}, we consider two cases that attack inputs $\mb{u}_a(t)$ does or does not affect the outputs $\check{\mb{y}}^{({\rm m})}(t)$ and consequently attack detectability in $\mb{r}_0(t)$. 

Zero-dynamics attack (ZDA) case: 
it can be shown that under condition \ref{cond:detection_condition_suff_0}, \eqref{eq:H_q_elements_kernel} holds (and so does \eqref{eq:H_q_elements_kernel_2_2}) only in the worst-case scenario, in the sense of attack detection, that none of the agents involved in intra-cluster switching links are affected by the ZDA in a safe mode. 
To this end, consider \eqref{eq:H_q_elements_kernel_relabeled} under which ZDA remains stealthy in residual $\mb{r}_0(t)$ in the safe modes and recall 
\begin{align}\label{eq:ZDA_traj}
    \mb{x}(t)=\bar{\mb{x}}(t)+ \tilde{\mb{x}}(t), \ \ \tilde{\mb{x}}_0e^{ \lambda_0 t},\ \ \forall\, t \in [t_0, +\infty),
\end{align}
in a stealthy ZDA case with $\tilde{\mb{x}}_0e^{ \lambda_0 t_1} \in \ker(\mb{C})$ being the initial condition of ZDA (cf. \eqref{eq:inv_zeros}, and \eqref{eq:zda_cond} in Proposition \ref{propo:stealthy_attacks}) at $t=t_1$ for a safe mode. 
Similar to \eqref{eq:obs_cent_INV_zero}, by evaluating ZDA condition \eqref{eq:inv_zeros} for the tuple $(\mb{A}_{\mb{q}}, \mb{B}, \mb{C})$ with $\mb{A}_{\mb{q}} = (\mb{A}_{1}+\Delta \mb{A}_{\mb{q}})$ and considering \eqref{eq:H_q_elements_2} we obtain
%
% {
% \small
\begin{equation}\label{eq:obs_cent_INV_zero_2}
    \begin{bmatrix}
        \lambda_0 I_{} - (\mb{A}_{1}-\mb{H}\mb{C}) 
        &
        (\mb{A}_{\mb{q}}-\mb{A}_{1}) 
        &
        -\mb{B} 
        \\ 
        \mb{C} & 0_{} & 0_{}
    \end{bmatrix}
    \begin{bmatrix}
        \tilde{\mb{e}}(t_1) \\ \tilde{\mb{x}}(t_1) \\ \mb{u}_a(t_1)
    \end{bmatrix}
    =
    \begin{bmatrix}
        \boldsymbol{0} \\ \boldsymbol{0} \\ \boldsymbol{0}
    \end{bmatrix},
\end{equation}
% \normalsize
% }
%
where as in \eqref{eq:obs_cent_INV_zero}, $\tilde{\mb{e}}(t_1) = \tilde{\mb{x}}(t_1) $ with $\tilde{\mb{x}}(t_1)=\tilde{\mb{x}}_0e^{ \lambda_0 t_1}$ and $ \mb{u}_a(t_1) = {\mb{u}}_0e^{ \lambda_0 t_1}$. Then \eqref{eq:obs_cent_INV_zero_2} is simplified to
\begin{equation}\label{eq:obs_cent_INV_zero_3}
    \begin{bmatrix}
        \lambda_0 I_{} - (\mb{A}_{\mb{q}}-\mb{H}\mb{C})
        % -\Delta \mb{A}_{\mb{q}} 
        & -\mb{B} \\ \mb{C} & 0_{}
    \end{bmatrix}
    \begin{bmatrix}
        \tilde{\mb{x}}(0) \\ \mb{u}_0
    \end{bmatrix}
    =
    \begin{bmatrix}
        \boldsymbol{0} \\ \boldsymbol{0}
    \end{bmatrix},
\end{equation}
where further simplification, similar to that in \eqref{eq:obs_cent_INV_zero}, and expanding it out yields
%
% \begin{align}\label{eq:}
%     \begin{bmatrix}
%         \lambda_0 I_{N} & -I_{N} & 0\\
%         \alpha (\Delta \Lc_{\mb{q}}) & 0_{N} &  0_{N} \\
%         \mb{C}_{\rm x}& 0 & 0_{}\\
%         0 & \mb{C}_{\rm v} & 0
%     \end{bmatrix}
%     %
%     \begin{bmatrix}
%         \tilde{x}(t_0) \\ \tilde{v}(t_0) \\ \mb{u}_0
%     \end{bmatrix}
%     %
%     =
%     %
%     \begin{bmatrix}
%         \boldsymbol{0} \\ \boldsymbol{0} \\ \boldsymbol{0}
%     \end{bmatrix}, 
% \end{align}
\begin{align}\label{eq:}
    \begin{bmatrix}
        \lambda_0 I_{N} & -I_{N} & 0\\
        \alpha (\Lc_{1} +  \Delta \Lc_{\mb{q}}) & (\lambda_0+\gamma) I_{N} &  -I_{\bar{\Fc}} \\
        \mb{C}_{\rm x}& 0 & 0_{}\\
        0 & \mb{C}_{\rm v} & 0
    \end{bmatrix}
    \begin{bmatrix}
        \tilde{x}(t_0) \\ \tilde{v}(t_0) \\ \mb{u}_0
    \end{bmatrix}
    =
    \begin{bmatrix}
        \boldsymbol{0} \\ \boldsymbol{0} \\ \boldsymbol{0}
    \end{bmatrix}, 
\end{align}
from which and also from \eqref{eq:zda_cond} we have
\begin{align}\label{eq:ZDA_ex_1}
    \lambda_0 \tilde{x}_i(t_0) &= \tilde{v}_i(t_0),\ \ \forall\, i \in \Vc,
    \\ \alpha \Lc_{1} \tilde{x}(t_0) + (\lambda_0+\gamma) \tilde{v}(t_0)- I_{\bar{\Fc}} \mb{u}_0 &\overset{\eqref{eq:zda_cond}}{=} \boldsymbol{0},
    \\ \label{eq:ZDA_ex_2}
    \Delta \Lc_{\mb{q}} \tilde{x}(t_0) &= \boldsymbol{0},
    \\ \label{eq:ZDA_ex_3}
    \mb{C}_{\rm x}\tilde{x}(t_0)=\boldsymbol{0},
    \ \
    \mb{C}_{\rm v}\tilde{v}(t_0)&=\boldsymbol{0}.
\end{align}
Then one can conclude from \eqref{eq:measurments}, \eqref{eq:ZDA_traj}, \eqref{eq:ZDA_ex_1}, and \eqref{eq:ZDA_ex_3}  that
\begin{align}\label{eq:ZDA_ex_4}
    \tilde{x}_i(t_0)=\tilde{v}_i(t_0) = 0 \implies \tilde{x}_i(t)=\tilde{v}_i(t) = 0 , \ \ \forall \, i \in \Mc \subset \Vc,
\end{align}
and by applying the same permutation as defined in \eqref{eq:graph_component_lap} and used in \eqref{eq:H_q_elements_kernel_relabeled} to equation \eqref{eq:ZDA_ex_2} as well as by considering \eqref{eq:ZDA_traj} and \eqref{eq:ZDA_ex_1} that
\begin{align}\label{eq:ZDA_ex_5}
    \tilde{x}_i(t_0)&=\tilde{x}_j(t_0) 
    \implies
    \tilde{x}_i(t)=\tilde{x}_j(t),
    \ \  \forall \, i,j \in \Dc_{\rm c} \subset \Vc, 
    \\ \label{eq:ZDA_ex_6}
    \tilde{v}_i(t_0)&=\tilde{v}_j(t_0) 
    \implies
    \tilde{v}_i(t)=\tilde{v}_j(t), 
    \ \ \hspace{.6ex}\forall \, i,j \in \Dc_{\rm c} \subset \Vc.
\end{align}
Also, as shown in Lemma \ref{lemma:app_component_measur}, under condition \ref{cond:detection_condition_suff_0} we have
\begin{align}\label{eq:graph_component_measurements}
    \Dc_{\rm c} \cap \Mc \neq \emptyset, \ \ \forall \, {\rm c} \in \{1, \cdots, \boldsymbol{{\rm c}}\},
\end{align}
with set $\Dc_{\rm c}$ given in \eqref{eq:states_relabeled}.

\noindent
Now under \eqref{eq:graph_component_measurements}, it is concluded from \eqref{eq:ZDA_ex_4}, \eqref{eq:ZDA_ex_5}-\eqref{eq:ZDA_ex_6} that
\begin{align}\label{eq:zda_simp}
    \tilde{x}_i(t)=\tilde{x}_j(t)=0, 
    \ \  
    \tilde{v}_i(t)=\tilde{v}_j(t)=0, 
    \ \ \forall \, i,j \in \Dc_{\rm c}, 
    %
    % \tilde{x}_i(t)=\tilde{x}_j(t)=0,
    % \ \ &\forall \, i,j \in \Dc_{\rm c} \subset \Vc, 
    % %
    % \\ \label{eq:}
    % \tilde{v}_i(t)=\tilde{v}_j(t)=0, 
    % \ \ &\forall \, i,j \in \Dc_{\rm c} \subset \Vc.
\end{align}
which by considering \eqref{eq:ZDA_traj} implies that \eqref{eq:H_q_elements_kernel_2_2} is simplified to
\begin{align}\label{eq:H_q_elements_kernel__}
   (\mathfrak{e}^{\top}_{i}-\mathfrak{e}^{\top}_{j}) {x}^{(\rm m)}(t) 
   &= \nonumber \\
   (\mathfrak{e}^{\top}_{i}-\mathfrak{e}^{\top}_{j}) \bar{x}^{(\rm m)}(t)
   &=
   0,
    &\forall& \, i, j \in \Dc_{\rm c}, \ \ \forall \,{\rm c} \in \{1, \cdots, \boldsymbol{\rm c}\}, 
    \nonumber \\
    & &\forall& \, {\rm m} \in \naturals_0, \ \ \forall \,t \in [t_1, +\infty),
\end{align}
where $\bar{x}_i$ and $\bar{x}_j$ are the elements of state vector $\bar{\mb{x}}$ in \eqref{eq:ZDA_traj} denoting the states of an attack-free system that satisfies \eqref{eq:stealthy} (i.e. $\dot{\bar{\mb{x}}} = \mb{A}_{\mb{q}} \bar{\mb{x}}$ obtained using $\dot{\mb{x}}$-dynamics in \eqref{eq:cl_sys} with $\mb{B}\mb{u}_a=\boldsymbol{0} $ and unknown initial condition $\bar{\mb{x}}_0$ as defined in Proposition \ref{propo:stealthy_attacks}).  
%
% it can be verified that system trajectories in \eqref{eq:H_q_elements_5} are the same as those of an attack-free dynamics satisfying \eqref{eq:stealthy} 
% (i.e. $\dot{\bar{\mb{x}}} = \mb{A}_{\mb{q}} \bar{\mb{x}}$ that is $\dot{\mb{x}}$-dynamics in \eqref{eq:cl_sys} with $\mb{B}\mb{u}_a=\boldsymbol{0} $ and unknown initial condition $\bar{\mb{x}}_0$ as defined in Proposition \ref{propo:stealthy_attacks}). 
%and thus for \eqref{eq:H_q_elements_5} we have
%
% \begin{equation}\label{eq:case2_outout}
%     \check{\mb{y}}^{({\rm m})} = \mathfrak{C} \mb{x}^{({\rm m})}= \mathfrak{C} \bar{\mb{x}}^{({\rm m})}=0, \, \forall \, {\rm m} \in \naturals_{0}, \, \forall \, t \in [t_1, \infty),
% \end{equation}
%
% \begin{align}\label{eq:}
%     (\mathfrak{e}^{\top}_{i}-\mathfrak{e}^{\top}_{j}) {x}^{(\rm m)}(t) = (\mathfrak{e}^{\top}_{i}-\mathfrak{e}^{\top}_{j}) \bar{x}^{(\rm m)}(t) =  \boldsymbol{0}, \, \forall \, i, j \in \Dc_{\rm i}, \, \forall \, {\rm m} \in \naturals_{0}, \, \forall \, t \in [t_1, \infty),
% \end{align}
%
% based on the equivalence relation in \eqref{eq:H_q_elements_kernel__}, we now 
Then using the attack-free dynamics $\dot{\bar{\mb{x}}} = \mb{A}_{\mb{q}} \bar{\mb{x}}$, the term $(\mathfrak{e}^{\top}_{i}-\mathfrak{e}^{\top}_{j}) \bar{x}^{(\rm m)}(t)
   =
   0$ in \eqref{eq:H_q_elements_kernel__} can be rewritten as
\begin{align}\label{eq:H_q_elements_new1}
    (\mathfrak{e}^{\top}_{i}-\mathfrak{e}^{\top}_{j}) \Lc^{\rm m}_{\mb{q}} \bar{{x}}(t) = 0, 
    \ \ \forall \, i, j \in \Dc_{\rm c},\, \forall \, {\rm m} \in \naturals_{0}, \, \forall \, t \in [t_1, \infty),
    \\ \label{eq:H_q_elements_new2}
    (\mathfrak{e}^{\top}_{i}-\mathfrak{e}^{\top}_{j}) \Lc^{\rm m}_{\mb{q}} \bar{{v}}(t) = 0,
    \ \  \, \forall \, i, j \in \Dc_{\rm c},\, \forall \, {\rm m} \in \naturals_{0}, \, \forall \, t \in [t_1, \infty).
\end{align}
Moreover, note that \eqref{eq:H_q_elements_new1} and \eqref{eq:H_q_elements_new2} have the same form as equations (109a) and  (109b) in \cite{mao2020novel}. Then under further conditions \ref{cond:detection_condition_suff_1} and \ref{cond:detection_condition_suff_2}, it can be verified using the same procedure as in \cite[Th. 2]{mao2020novel} that \eqref{eq:H_q_elements_new1} and \eqref{eq:H_q_elements_new2} yield
\begin{align}\label{eq:equib_1}
    \bar{x}_i^{}(t) = \bar{x}_j^{}(t), \ \ 
    \forall \, i, j \in \Vc, \ \ \forall \, t \in [t_1, +\infty),
    \\ \label{eq:equib_2}
    \bar{v}_i^{}(t) = \bar{v}_j^{}(t), \ \
    \forall \, i, j \in \Vc, \ \ \forall \, t \in [t_1, +\infty),
\end{align}
which means the the entire states of the attack-free system have achieved consensus. Considering the equilibrium subspace \eqref{eq:cond_consensus} as a result of the consensus protocol \eqref{eq:ctrl_proto}, one can conclude that \eqref{eq:equib_1}-\eqref{eq:equib_2} and \eqref{eq:cond_consensus} coincide. Therefore, from \eqref{eq:H_q_elements_kernel__} and \eqref{eq:equib_1}-\eqref{eq:equib_2}, obtained under conditions \ref{cond:detection_condition_suff_0}-\ref{cond:detection_condition_suff_2}, one can conclude that stealthy ZDA is undetectable in $\mb{r}_0(t)$ of \eqref{eq:obs_cent} only in the worst-case scenario that intra-cluster switching links are between agents whose trajectories are not affected by ZDA as well as all of the system \eqref{eq:cl_sys}'s attack-free trajectories, characterized in \eqref{eq:H_q_elements_kernel__}, are at the consensus equilibrium \eqref{eq:cond_consensus}.

Covert attack case:
consider \eqref{eq:H_q_elements_kernel_relabeled} under which a covert attack remains stealthy in a safe mode and note that
% $\mb{x}(t)=\bar{\mb{x}}(t)+\tilde{\mb{x}}(t), \forall\, t \in [t_1, +\infty)$ with 
%
\begin{align}\label{eq:covert_sol}
\mb{x}(t)&=\bar{\mb{x}}(t)+\tilde{\mb{x}}(t),\ \
\forall\, t \in [t_1, +\infty), \;\; \text{with}
\nonumber \\
  \tilde{\mb{x}}(t) &= 
	{e^{\mb{A}_1(t-t_{1})}}
	\tilde{\mb{x}}(t_1)  
	+
	{\int^{t}_{t_{1}}
	e^{\mb{A}_{1}(t-\boldsymbol{\tau})}
	\mb{B}\mb{u}_{a}(\boldsymbol{\tau})d{\boldsymbol{\tau}}}
\end{align}
according to the attack model \eqref{eq:attack_model} and Proposition \ref{propo:stealthy_attacks}.
Given \eqref{eq:covert_sol}, \eqref{eq:H_q_elements_kernel_2_2} can be rewritten as
\begin{align}\label{eq:H_q_elements_kernel_2_3}
    [(\mathfrak{e}^{\top}_{i}-\mathfrak{e}^{\top}_{j}) \ \ 0]\, \bar{\mb{x}}^{(\rm m)}(t) 
    =
    [(\mathfrak{e}^{\top}_{i}-\mathfrak{e}^{\top}_{j}) \ \ 0]\, \tilde{\mb{x}}^{(\rm m)}(t), \ \ &\forall \, i, j \in \Dc_{\rm c}, 
    \nonumber \\
   \forall \, {\rm c} \in \{1, \cdots, \boldsymbol{\rm c}\},  \ \
    \forall \, {\rm m} \in \naturals_0, \ \ \forall \,t \in [t_1, +\infty),
\end{align}
Notice that the attack-free system states, $\bar{\mb{x}}(t)$ in \eqref{eq:covert_sol}, converge to \eqref{eq:cond_consensus} as $t\rightarrow +\infty$, then the left side of   \eqref{eq:H_q_elements_kernel_2_3} converges to zero and one can conclude from \eqref{eq:covert_sol} and \eqref{eq:H_q_elements_kernel_2_3} that continuous states $\tilde{\mb{x}}(t) = \col(\tilde{x}(t),\tilde{v}(t))$ exist in either of the following cases
\begin{align}\label{eq:covert_case1}
\hspace{-1ex}\text{case 1}:
\tilde{{x}}_{i}(t)=\tilde{{x}}_{j}(t)  \neq 0, 
\ \ &\forall \, i, j \in \Dc_{\rm c}, \, \forall\, t \in [t_1, +\infty)
\end{align}
\begin{align}\label{eq:eq:covert_case2}
\hspace{-1ex}\text{case 2}:
\tilde{{x}}_{i}(t)=\tilde{{x}}_{j}(t) = 0, 
\ \ &\forall \, i, j \in \Dc_{\rm c}, \, \forall\, t \in [t_1, +\infty)
\end{align}
Note that here case 1 in \eqref{eq:covert_case1} implies the attack input $\mb{u}_{a}$ in \eqref{eq:covert_sol} has driven and kept the states of agents involved in switching into an unknown equilibrium over time span $\forall \, t \in [t_1, +\infty)$. Also, case 2's interpretation and analysis coincide with that of ZDA in \eqref{eq:zda_simp}. Then following the same analysis as the ZDA's, one can conclude that, under conditions \ref{cond:detection_condition_suff_0}-\ref{cond:detection_condition_suff_2}, covert attack is undetectable in $\mb{r}_0(t)$ of \eqref{eq:obs_cent} only in the worst-case scenarios that 1) intra-cluster switching links are between agents whose trajectories are identical over time under the effect of covert attack; and 2) intra-cluster switching links are between agents whose trajectories are not affected by covert attack as well as all of the system \eqref{eq:cl_sys}'s attack-free trajectories are at the consensus equilibrium \eqref{eq:cond_consensus}.

\bibliographystyle{IEEEtran}
\bibliography{IEEEabrv,references.bib}

\end{document}

%% file: preamble.tex
\usepackage{import}
\usepackage{graphics}
\usepackage{graphicx}
\usepackage{amssymb,amsmath,mathrsfs}
\usepackage{mathtools}
\usepackage{stackengine}

\usepackage{color,xcolor}
\definecolor{p2color}{HTML}{D95319}
\definecolor{pscolor}{HTML}{77AC30}
\definecolor{hpscolor}{HTML}{7E2F8E}

\usepackage[font=small,caption=false]{subfig}
\usepackage{algorithm}
\usepackage{algpseudocode} 
\usepackage{etex}
\usepackage{nicematrix}
\usepackage{arydshln}
\usepackage{dashrule}

\usepackage[normalem]{ulem}

\usepackage{multirow}
\usepackage{booktabs}

\usepackage{enumitem}
\renewcommand{\theenumi}{\rm{(\roman{enumi})}}

\usepackage{amsthm}
\newcommand{\btitle}[1]{\mbox{}{\bf{(#1).}}}
\newtheorem{theorem}{Theorem}[section]
\newtheorem{remark}{Remark}[theorem]

\newtheorem{assumption}{Assumption}
\newtheorem{proposition}[theorem]{Proposition}
\newtheorem{lemma}[theorem]{Lemma}
\newtheorem{definition}[theorem]{Definition}

\newtheorem{problem}{Problem}

\newcommand{\real}{\ensuremath{\mathbb{R}}}

\newcommand{\naturals}{\ensuremath{\mathbb{N}}}
\newcommand{\cplx}{\ensuremath{\mathbb{C}}}

\newcommand{\mb}[1]{\mathbf{ #1 }}

\newcommand{\supp}{\ensuremath{\mathrm{supp}}}
\newcommand{\colsupp}{\ensuremath{\mathrm{colsupp}}}

\newcommand{\col}{\mathrm{col}}
\newcommand{\rank}{\mathrm{rank}}
\newcommand{\spann}{\mathrm{span}}
\newcommand{\diag}{\mathrm{diag}}

\newcommand{\Img}{\mathrm{Im}}

\newcommand{\setdef}[2]{\left\{#1 \mid #2\right\}}
\newcommand{\paren}[1]{\left(#1\right)}
\newcommand{\bracket}[1]{\left[#1\right]}
\newcommand{\braces}[1]{\left\{#1\right\}}

\newcommand{\s}{\scriptscriptstyle}

\newcommand{\spi}{\scriptscriptstyle \mathcal{P}_{\!i}}

\newcommand{\sigt}{\sigma(t)}

\newcommand{\Ac}{\mathcal{A}}

\newcommand{\Dc}{\mathcal{D}}
\newcommand{\Ec}{\mathcal{E}}
\newcommand{\Fc}{\mathcal{F}}
\newcommand{\Gc}{\mathcal{G}}
\newcommand{\Hc}{\mathcal{H}}

\newcommand{\Lc}{\mathcal{L}}
\newcommand{\Mc}{\mathcal{M}}
\newcommand{\Nc}{\mathcal{N}}
\newcommand{\Oc}{\mathcal{O}}
\newcommand{\Pc}{\mathcal{P}}
\newcommand{\Qc}{\mathcal{Q}}

\newcommand{\Uc}{\mathcal{U}}
\newcommand{\Vc}{\mathcal{V}}

\newcommand{\Xc}{\mathcal{X}}
\newcommand{\Yc}{\mathcal{Y}}
\newcommand{\Zc}{\mathcal{Z}}

\usepackage[normalem]{ulem}

%% file: main.bbl
\begin{thebibliography}{10}
\providecommand{\url}[1]{#1}
\csname url@samestyle\endcsname
\providecommand{\newblock}{\relax}
\providecommand{\bibinfo}[2]{#2}
\providecommand{\BIBentrySTDinterwordspacing}{\spaceskip=0pt\relax}
\providecommand{\BIBentryALTinterwordstretchfactor}{4}
\providecommand{\BIBentryALTinterwordspacing}{\spaceskip=\fontdimen2\font plus
\BIBentryALTinterwordstretchfactor\fontdimen3\font minus
  \fontdimen4\font\relax}
\providecommand{\BIBforeignlanguage}[2]{{%
\expandafter\ifx\csname l@#1\endcsname\relax
\typeout{** WARNING: IEEEtran.bst: No hyphenation pattern has been}%
\typeout{** loaded for the language `#1'. Using the pattern for}%
\typeout{** the default language instead.}%
\else
\language=\csname l@#1\endcsname
\fi
#2}}
\providecommand{\BIBdecl}{\relax}
\BIBdecl

\bibitem{cardenas2008secure}
A.~A. Cardenas, S.~Amin, and S.~Sastry, ``Secure control: Towards survivable
  cyber-physical systems,'' in \emph{2008 The 28th International Conference on
  Distributed Computing Systems Workshops}.\hskip 1em plus 0.5em minus
  0.4em\relax IEEE, 2008, pp. 495--500.

\bibitem{pasqualetti2013attack}
F.~Pasqualetti, F.~D{\"o}rfler, and F.~Bullo, ``Attack detection and
  identification in cyber-physical systems,'' \emph{IEEE transactions on
  automatic control}, vol.~58, no.~11, pp. 2715--2729, 2013.

\bibitem{ren2007distributed}
W.~Ren and E.~Atkins, ``Distributed multi-vehicle coordinated control via local
  information exchange,'' \emph{International Journal of Robust and Nonlinear
  Control: IFAC-Affiliated Journal}, vol.~17, no. 10-11, pp. 1002--1033, 2007.

\bibitem{pasqualetti2015divide}
F.~Pasqualetti, F.~D{\"o}rfler, and F.~Bullo, ``A divide-and-conquer approach
  to distributed attack identification,'' in \emph{2015 54th IEEE Conference on
  Decision and Control (CDC)}.\hskip 1em plus 0.5em minus 0.4em\relax IEEE,
  2015, pp. 5801--5807.

\bibitem{teixeira2012revealing}
A.~Teixeira, I.~Shames, H.~Sandberg, and K.~H. Johansson, ``Revealing stealthy
  attacks in control systems,'' in \emph{2012 50th Annual Allerton Conference
  on Communication, Control, and Computing (Allerton)}.\hskip 1em plus 0.5em
  minus 0.4em\relax IEEE, 2012, pp. 1806--1813.

\bibitem{hoehn2016detection}
A.~Hoehn and P.~Zhang, ``Detection of covert attacks and zero dynamics attacks
  in cyber-physical systems,'' in \emph{2016 American Control Conference
  (ACC)}.\hskip 1em plus 0.5em minus 0.4em\relax IEEE, 2016, pp. 302--307.

\bibitem{schellenberger2017detection}
C.~Schellenberger and P.~Zhang, ``Detection of covert attacks on cyber-physical
  systems by extending the system dynamics with an auxiliary system,'' in
  \emph{2017 IEEE 56th Annual Conference on Decision and Control (CDC)}.\hskip
  1em plus 0.5em minus 0.4em\relax IEEE, 2017, pp. 1374--1379.

\bibitem{mao2020novel}
Y.~Mao, H.~Jafarnejadsani, P.~Zhao, E.~Akyol, and N.~Hovakimyan, ``Novel
  stealthy attack and defense strategies for networked control systems,''
  \emph{IEEE Transactions on Automatic Control}, 2020.

\bibitem{barboni2020detection}
A.~Barboni, H.~Rezaee, F.~Boem, and T.~Parisini, ``Detection of covert
  cyber-attacks in interconnected systems: A distributed model-based
  approach,'' \emph{IEEE Transactions on Automatic Control}, 2020.

\bibitem{jafarnejadsani2018multirate}
H.~Jafarnejadsani, H.~Lee, N.~Hovakimyan, and P.~Voulgaris, ``A multirate
  adaptive control for mimo systems with application to cyber-physical
  security,'' in \emph{2018 IEEE Conference on Decision and Control
  (CDC)}.\hskip 1em plus 0.5em minus 0.4em\relax IEEE, 2018, pp. 6620--6625.

\bibitem{back2017enhancement}
J.~Back, J.~Kim, C.~Lee, G.~Park, and H.~Shim, ``Enhancement of security
  against zero dynamics attack via generalized hold,'' in \emph{2017 IEEE 56th
  Annual Conference on Decision and Control (CDC)}.\hskip 1em plus 0.5em minus
  0.4em\relax IEEE, 2017, pp. 1350--1355.

\bibitem{sundaram2010distributed}
S.~Sundaram and C.~N. Hadjicostis, ``Distributed function calculation via
  linear iterative strategies in the presence of malicious agents,'' \emph{IEEE
  Transactions on Automatic Control}, vol.~56, no.~7, pp. 1495--1508, 2010.

\bibitem{mitra2016secure}
A.~Mitra and S.~Sundaram, ``Secure distributed observers for a class of linear
  time invariant systems in the presence of byzantine adversaries,'' in
  \emph{2016 IEEE 55th Conference on Decision and Control (CDC)}.\hskip 1em
  plus 0.5em minus 0.4em\relax IEEE, 2016, pp. 2709--2714.

\bibitem{teixeira2010networked}
A.~Teixeira, H.~Sandberg, and K.~H. Johansson, ``Networked control systems
  under cyber attacks with applications to power networks,'' in
  \emph{Proceedings of the 2010 American Control Conference}.\hskip 1em plus
  0.5em minus 0.4em\relax IEEE, 2010, pp. 3690--3696.

\bibitem{gallo2020distributed}
A.~J. Gallo, M.~S. Turan, F.~Boem, T.~Parisini, and G.~Ferrari-Trecate, ``A
  distributed cyber-attack detection scheme with application to dc
  microgrids,'' \emph{IEEE Transactions on Automatic Control}, vol.~65, no.~9,
  pp. 3800--3815, 2020.

\bibitem{anguluri2018attack}
R.~Anguluri, V.~Katewa, and F.~Pasqualetti, ``Attack detection in stochastic
  interconnected systems: Centralized vs decentralized detectors,'' in
  \emph{2018 IEEE Conference on Decision and Control (CDC)}.\hskip 1em plus
  0.5em minus 0.4em\relax IEEE, 2018, pp. 4541--4546.

\bibitem{newman2018networks}
M.~Newman, \emph{Networks}.\hskip 1em plus 0.5em minus 0.4em\relax Oxford
  university press, 2018.

\bibitem{zhou1996robust}
K.~Zhou, J.~C. Doyle, K.~Glover \emph{et~al.}, \emph{Robust and optimal
  control}.\hskip 1em plus 0.5em minus 0.4em\relax Prentice hall New Jersey,
  1996, vol.~40.

\bibitem{tanwani2012observability}
A.~Tanwani, H.~Shim, and D.~Liberzon, ``Observability for switched linear
  systems: characterization and observer design,'' \emph{IEEE Transactions on
  Automatic Control}, vol.~58, no.~4, pp. 891--904, 2012.

\bibitem{chen1996design}
J.~Chen, R.~J. Patton, and H.-Y. Zhang, ``Design of unknown input observers and
  robust fault detection filters,'' \emph{International Journal of control},
  vol.~63, no.~1, pp. 85--105, 1996.

\bibitem{chesi2011nonconservative}
G.~Chesi, P.~Colaneri, J.~C. Geromel, R.~Middleton, and R.~Shorten, ``A
  nonconservative lmi condition for stability of switched systems with
  guaranteed dwell time,'' \emph{IEEE Transactions on Automatic Control},
  vol.~57, no.~5, pp. 1297--1302, 2011.

\bibitem{teixeira2014distributed}
A.~Teixeira, I.~Shames, H.~Sandberg, and K.~H. Johansson, ``Distributed fault
  detection and isolation resilient to network model uncertainties,''
  \emph{IEEE transactions on cybernetics}, vol.~44, no.~11, pp. 2024--2037,
  2014.

\bibitem{chilali1996h}
M.~Chilali and P.~Gahinet, ``H/sub/spl infin//design with pole placement
  constraints: an lmi approach,'' \emph{IEEE Transactions on automatic
  control}, vol.~41, no.~3, pp. 358--367, 1996.

\bibitem{chen2004observer}
W.~Chen and S.~Mehrdad, ``Observer design for linear switched control
  systems,'' in \emph{Proceedings of the 2004 American Control Conference},
  vol.~6.\hskip 1em plus 0.5em minus 0.4em\relax IEEE, 2004, pp. 5796--5801.

\bibitem{meng2018consensus}
H.~Meng, Z.~Chen, and R.~Middleton, ``Consensus of multiagents in switching
  networks using input-to-state stability of switched systems,'' \emph{IEEE
  Transactions on Automatic Control}, vol.~63, no.~11, pp. 3964--3971, 2018.

\bibitem{bernstein2009matrix}
D.~S. Bernstein, \emph{Matrix mathematics}.\hskip 1em plus 0.5em minus
  0.4em\relax Princeton university press, 2009.

\bibitem{olfati2004consensus}
R.~Olfati-Saber and R.~M. Murray, ``Consensus problems in networks of agents
  with switching topology and time-delays,'' \emph{IEEE Transactions on
  automatic control}, vol.~49, no.~9, pp. 1520--1533, 2004.

\bibitem{hespanha2018linear}
J.~P. Hespanha, \emph{Linear systems theory}.\hskip 1em plus 0.5em minus
  0.4em\relax Princeton university press, 2018.

\bibitem{lee2017l1}
H.~Lee, ``L1 adaptive control for nonlinear and non-square multivariable
  systems,'' Ph.D. dissertation, University of Illinois at Urbana-Champaign,
  2017.

\end{thebibliography}
